\documentclass[aps,a4paper,superscriptaddress,twocolumn,preprintnumbers]{revtex4-1}

\usepackage[german,english]{babel}
\usepackage{amsmath}    
\usepackage[colorlinks=true,citecolor=black,linkcolor=black]{hyperref}

\usepackage{amsfonts}
\usepackage{amssymb}
\usepackage{bm}
\usepackage{bbold}

\usepackage[T1]{fontenc}
\usepackage{times}
\usepackage{fix-cm}

\usepackage{natbib}
\usepackage{graphicx}

\usepackage[labelfont=footnotesize,font=footnotesize]{caption} 
\usepackage[labelfont=scriptsize,font=scriptsize,singlelinecheck=off,
justification=centering]{subcaption}
\usepackage{booktabs} 
\usepackage{tabularx}
\usepackage{tabulary}
\newcolumntype{Z}{>{\centering\arraybackslash}X} 

\newcommand{\ket}[1]{\mathop{\left|#1\right\rangle}}
\newcommand{\braket}[2]{\langle #1 | #2 \rangle}

\newcommand{\pouter}[2]{\ensuremath{\mathopen{}\left|{#1}\right\rangle\!\left\langle{#2}\right|}} 
\newcommand{\idd}{\id}
\newcommand{\id}{\mathbb{1}}
\newcommand{\bpm}{\begin{pmatrix}}
\newcommand{\epm}{\end{pmatrix}}

\newcommand{\mbf}[1]{\mathbf{#1}}

\newcommand{\I}{\mathrm{i}}
\newcommand{\II}{\mathrm{i}}

\newcommand{\MTR}{\text{Tr}}
\newcommand{\mtr}{\MTR}
\newcommand{\hilb}{\mathcal{H}}
\newcommand{\mink}{\mathbf{M}}
\newcommand{\RRR}{\mathbb{R}^3}
\newcommand{\CC}{\mathbb{C}^2}
\newcommand{\sun}{\ensuremath{\mathrm{SU}(N)}}

\newcommand{\SNW}{\mathbf{S}_{\mathrm{NW}}}
\newcommand{\plane}[2]{$#1#2$\nobreakdash-plane}

\newcommand{\dd}[1]{\mathrm{d}#1}
\newcommand{\dddd}{\mathrm{d}}

\newcommand{\tum}[1]{\,\mathrm{d}\mu(#1)}
\newcommand{\mut}[1]{\mathrm{d}\mu(#1)\,}
\newcommand{\uuu}[1]{\mathrm{d}\mu(#1)} 
\newcommand{\ppp}{\mathbf{p}}
\newcommand{\qqq}{\mathbf{q}}

\newcommand{\vvv}{\mathbf{v}}
\newcommand{\nnn}{\mathbf{n}}
\newcommand{\SSS}{\mathbf{S}}
\newcommand{\WWW}{\mathbf{W}}
\newcommand{\sssigma}{\boldsymbol\sigma}
\newcommand{\pppp}{p}
\newcommand{\qqqq}{q}

\newcommand{\degree}{\ensuremath{^\circ}}
\newcommand{\comment}[1]{}

\begin{document}

\title{Relativistic entanglement of two particles driven by continuous product
momenta}
\author{Veiko Palge}
\email[Email: ]{veiko.palge@ut.ee}
\affiliation{Laboratory of Theoretical Physics, Institute of Physics,
University of Tartu, W. Ostwaldi 1, 50411 Tartu, Estonia}

\author{Jacob Dunningham}
\email[Email: ]{J.Dunningham@sussex.ac.uk}
\affiliation{Department of Physics and Astronomy, University of Sussex,
Brighton BN1 9QH, United Kingdom}

\author{Stefan Groote}
\email[Email: ]{stefan.groote@ut.ee}
\affiliation{Laboratory of Theoretical Physics, Institute of Physics,
University of Tartu, W. Ostwaldi 1, 50411 Tartu, Estonia}

\author{Hannes Liivat}
\email[Email: ]{hannes.liivat@ut.ee}
\affiliation{Laboratory of Theoretical Physics, Institute of Physics,
University of Tartu, W. Ostwaldi 1, 50411 Tartu, Estonia}

\begin{abstract}
In this paper we explore the entanglement of two relativistic spin-$1/2$
particles with continuous momenta. The spin state is described by the Bell
state and the momenta are given by Gaussian distributions of product form.
Transformations of the spins are systematically investigated in different
boost scenarios by calculating the orbits and concurrence of the spin degree
of freedom. By visualizing the behavior of the spin state we get further
insight into how and why the entanglement changes in different boost
situations.
\end{abstract}

\maketitle

\section{Introduction}

Entanglement is the key notion that distinguishes between the quantum and
classical world. It has also proven extremely useful for applications in the
context of quantum information theory. While most of the theory of entanglement
is non-relativistic, the ultimate description of reality is given by the
relativistic theory, thus a complete account of entanglement demands that we
understand how entanglement behaves in relativity.

The field of relativistic quantum information, where the first studies appeared
a little more than a decade ago, is an attempt to provide such an account
\cite{czachor_einstein-podolsky-rosen-bohm_1997, gingrich_quantum_2002,
peres_quantum_2002, alsing_lorentz_2002, ahn_relativistic_2002,
ahn_relativistic_2003, ahn_lorentz-covariant_2003, czachor_relativistic_2003,
moon_relativistic_2004, lee_quantum_2004, czachor_comment_2005,
lamata_relativity_2006, alsing_entanglement_2006, chakrabarti_entangled_2009,
fuentes_entanglement_2010, friis_relativistic_2010, palge_generation_2012}.
The overall conclusion emerging from this work is that relativistic entanglement
in both inertial and accelerated frames is observer dependent
\cite{alsing_observer-dependent_2012}. The issue has been in the spotlight since
early on. It was found in \cite{gingrich_quantum_2002} that the entanglement of
a Bell state generally decreases under Lorentz boosts. Almost simultaneously, it
was reported in \cite{alsing_lorentz_2002} that although boosted particles
undergo Wigner rotations, the entanglement fidelity of a Bell state remains
invariant. This resulted in a number of followup papers, see e.g.\
\cite{terashima_relativistic_2003, caban_lorentz-covariant_2005,
caban_einstein-podolsky-rosen_2006, jordan_maps_2006, jordan_lorentz_2007}, some
of which confirm the invariance of entanglement while others claim that
entanglement depends on the boost in question. On closer inspection one notices
that the (sometimes seemingly contradictory) results in the literature rely on
different momentum states and boost angles, or geometries, involved.
That geometry plays an instrumental role in determining the behavior of
entanglement under Lorentz boosts is also suggested by a study of the simplest
system, the single particle \cite{palge_generation_2012}.  Likewise the
literature on the Wigner rotation is quite clear about the fact that its nature
is highly geometric, yet barring a few cases \cite{palge_behavior_2011}, there
is almost no work in relativistic quantum information that systematically takes
this into account.

In this paper, we aim to fill the lacuna by exploring a number of boost
situations with different momenta as well as geometries. The focus is on
massive two particle spin-$1/2$ systems whose momentum states are given by
continuous distributions of product form \footnote{A related study that
involves systems with discrete momenta along with an extension to mixed spin
states can be found in \cite{palge_behavior_2015}.}. We assume that the spin
degree of freedom is described by a maximally entangled Bell state. We
visualize the spin state in a 3D manner in order to gain a better understanding
of how and why entanglement changes under boosts. The aim of the paper is to
provide a simple model that helps explain the various results obtained so far
for systems with continuous momenta. Another is to contribute to a survey of
different momentum states and geometries in order to have a better view of the
landscape of systems that might be of interest for relativistic quantum
information theory.

The paper is organized as follows. We begin by outlining how a generic two
particle state transforms under Lorentz boosts. The properties of Wigner
rotation will then be reviewed, followed by a specification of the models we
will study below. Sections \ref{sec:momentaAndSpinRotations} and
\ref{sec:spinStateAndItsVisualization} give a detailed characterization of the
momentum and spin states of the models, respectively. The second half of the
paper from section \ref{sec:productMomenta} onwards examines how spin
entanglement changes in two particle systems that contain various forms of
product momenta. We summarize the results in section \ref{sec:conclusion}.

\section{General setting}

In this paper, we will study a system consisting of two massive spin-$1/2$
particles and ask how the entanglement of spins changes when viewed from a
different inertial frame. This question is uninteresting in the
non-relativistic setting because boosts do not change the spin state. However,
in the relativistic world the situation is non-trivial. The spin seen by an
observer in any other frame generally depends on the momentum of the particle
and the state of the observer. This entails that spin entanglement in general
changes non-trivially too. We begin the discussion by fixing the state space
and calculating the generic transformation of a two particle state under
Lorentz transformations.

Free spin-$1/2$ particles can be described in two different theories, the
unitary irreducible representation of the Poincar{\'e} group or in the Dirac
theory of bispinors \footnote{The unitary irreducible representation is due to
Wigner \cite{wigner_unitary_1939}. A standard treatment of the two
representations is given by \cite{bogolubov_introduction_1975}. See
\cite{polyzou_spin_2012} for a very readable account on the relationship
between the two frameworks and \cite{caban_spin_2013} for a discussion in the
context of spin observable in the Dirac theory.}. Throughout we will work in
the Wigner representation (also called the Wigner-Bargmann or the spin basis
\cite{caban_spin_2013}) and use basis vectors of the form $\ket{\ppp, \lambda}
\equiv \ket{\ppp}\ket{\lambda} \equiv \ket{\ppp} \otimes \ket{\lambda}$, where
$\ppp$ labels the single particle momentum and $\lambda = \pm \frac{1}{2}$ is
the spin (see Appendix~\ref{sec:AppendixA} for constructions used in the
paper). A generic pure two particle state $\ket{\Psi} \in \hilb \otimes \hilb$
can be written as follows,
\begin{align}\label{eq:genericStateTwoParticlesContinuousMomentum}
\ket{\Psi} = \sum_{\lambda\eta} \int \mut{\pppp, \qqqq} \psi_{\lambda\eta}(\ppp,
\qqq) \ket{\ppp, \lambda}\ket{\qqq, \eta},
\end{align}
where $\uuu{\pppp} = \left[2E(\ppp)\right]^{-1}\dd{\ppp}$ is the Lorentz
invariant integration measure, we have abbreviated $\uuu{\pppp, \qqqq} =
\uuu{\pppp}\uuu{\qqqq}$ and $\hilb = L^2(\RRR) \otimes \CC$ denotes the single
particle states space. The wave function satisfies the normalization condition
\begin{align}\label{eq:relativisticNormalization}
\sum_{\lambda\eta} \int \tum{\pppp, \qqqq} |\psi_{\lambda\eta}(\ppp, \qqq)|^2
= 1,
\end{align}
and the (improper) spin and momentum eigenstates satisfy the orthogonality
condition
\begin{align}\label{eq:theOrthogonalityConditionImproperSpinMomentum}
\langle \ppp', \lambda' | \ppp, \lambda \rangle = 2E(\ppp) \delta^{3}(\ppp -
\ppp') \delta_{\lambda\lambda'}.
\end{align}
An observer $O^{\Lambda}$ who is Lorentz boosted relative to $O$ by $\Lambda$ describes the same system using a different wave function
\begin{align}\label{eq:unitaryActionOnFiberTwoParticles}
\psi^{\Lambda}_{\kappa\nu}(\ppp, \qqq) &= \sum_{\lambda\eta}
D_{\kappa\lambda}\left[W(\Lambda, \Lambda^{-1}\ppp)\right]
D_{\nu\eta}\left[W(\Lambda, \Lambda^{-1}\qqq)\right] \nonumber\\
&\phantom{\sum}\times\psi_{\lambda\eta}(\Lambda^{-1}\ppp, \Lambda^{-1}\qqq),
\end{align}
where $D\left[W(\Lambda, \ppp)\right] \in \mathrm{SU}(2)$ is the spin-$1/2$
representation of the Wigner, or Thomas--Wigner rotation (TWR), $W =
L^{-1}(\Lambda L_p) \Lambda L_p$ \footnote{We will use the abbreviation TWR in
honor of Thomas's contribution of discovering the Thomas precession
\cite{thomas_motion_1926,thomas_kinematics_1927}.}. This entails that for
$O^\Lambda$ the spins are rotated by $D[W(\Lambda, \ppp)]$ and the rotation
depends on the geometry, i.e.\ the angle between two boosts as well as the the
momenta of the system and the observer. Note that since each spin undergoes a
momentum dependent rotation, the result is a non-trivial transformation on the
spin degree of freedom of the total two particle state. This implies that
properties like entanglement which are defined in terms of spin will change in
general as well.

From the logical point of view, we can think of the two particle system as made
up of two spin qubits, where each spin qubit is controlled by a momentum system
\cite{peres_quantum_2002, gingrich_quantum_2002}. The analogy is from quantum
information theory where a controlled unitary gate consists of two input qubits
which are called the control qubit and the target qubit. The action of the gate
is to transform the target qubit with a unitary transformation $U$ depending on
the control qubit. One can conceive of the Lorentz boost along the same lines
\footnote{With the caveat that strictly considered the momentum state is not a
qubit, so we call it a momentum system.}. If momentum takes the role of a
control system, then given that the boost angle and rapidity are fixed, the
transform on the spin state depends only on the momentum state. While the idea
will not enter calculations, the notion of Lorentz boosts as controlled
unitaries will guide our investigation of the relativistic spin--momentum
systems in this paper. It prompts us to ask the question of what are the maps
that different momentum states generate on the spin degree of freedom? This is
a broad question and we will not try address it in a single paper. We will
approach the topic step-by-step by exploring how a particular subset of
interesting momenta drives the spin entanglement. In this paper, we will focus
on momenta that are of product form and ask what kind of transformations they
induce on the maximally entangled spin state of a two particle system
\footnote{Entangled momenta are discussed in a related paper.}? Further, while
previous work has investigated systems with discrete momenta
\cite{palge_behavior_2015}, which represent idealized models, realistic
situations are described by states whose momenta are given by continuous
distributions. To understand how the behavior of entanglement is affected when
the idealization is dropped we will assume that momenta are given by entangled
states that consist of combinations of Gaussians.

\section{Spin observable}

In contrast to the non-relativistic theory, treatment of spin in relativity
requires some care. This is due to the fact that the commutation relation of
two generators of rotationless Lorentz boosts results in a rotation generator,
$[K_i, N_j] = -i \epsilon_{ijk} J_k$. The latter is the infinitesimal algebraic
form of the TWR. It means that two non-collinear rotationless Lorentz boosts
will generate a rotation. From the geometric point of view it is interesting to
note that the same phenomenon is related to the fact that the relativistic
momentum space, the mass shell hyperbola, is a curved space: a Riemannian space
with constant negative curvature \cite{sexl_relativity_2001}.

While there is some controversy about what is the most adequate spin operator
in the relativistic quantum theory, one candidate stands out. It is the
so-called Newton-Wigner spin observable, which has advantages over other spins
because it possesses a number of properties one naturally demands of a good
spin operator. We will give a brief summary of the reasoning that leads to the
Newton-Wigner spin. A good overview along with the discussion of the various
spin candidates can be found in \cite{caban_spin_2013}.

Relativistic quantum theory conceptualizes particles as group representations.
Elementary particles correspond to the unitary irreducible representations of
the Poincar{\'e} group, which are characterized by two labels, mass $m$ and
spin $s$. Mass is given by the square root of the eigenvalues of the first
Casimir invariant, the mass square operator $P^2 = P_\mu P^\mu$. Spin is
related to the eigevalues of the second Casimir invariant, $W^2 = W_\mu W^\mu$,
where
\begin{align}
W^\mu = \frac{1}{2} \epsilon^{\nu \alpha \beta \mu} P_\nu J_{\alpha\beta}
\end{align}
is the Pauli-Lubanski vector and $J_{\alpha\beta}$ are the generators of the
Lorentz group.
The components of $W = (W^0, W^j)$ are given by
\begin{align}
W^0 &= P^j M^j = \mbf{P} \cdot \mbf{M}, \nonumber\\
W^j &= P^0 M^j - \epsilon^{jkl} P^k
N^l = P^0 M^j - (\mbf{P} \times \mbf{N})^j.\label{eq:PauliLubanskiVector}
\end{align}
One can then define the spin square operator as
\begin{align}
\SSS^2 = -\frac{1}{m^2} W_\mu W^\mu.
\end{align}
This leads to the idea that the spin operator can be postulated as a linear
combination of the components of $W$ given that certain conditions are
satisifed, conditions that one would reasonably require of a spin observable.
These are as follows, (i)~the spin operator $\SSS$ should fulfill the usual
commutation relations,
\begin{align}
[S^i, S^j] = \I \epsilon_{ijk} S^k,
\end{align}
(ii) it is a three dimensional vector, that is
\begin{align}
[J^i, S^j] = \I \epsilon_{ijk} S^k,
\end{align}
and (iii) in any frame the vector $\SSS$ is a linear combination of components
of $W$ with coefficients that depend only on the four momentum $P^\mu$. It can
be shown that there is a unique linear combination of operators $W^\mu$ which
satisfies these conditions and it has the form
\cite{bogolubov_introduction_1975},
\begin{align}
\SNW = \frac{1}{m}
\left(\WWW - \frac{W_0}{m + P^0}\mbf{P} \right).\label{eq:NewtonWignerS}
\end{align}
The Newton-Wigner observable corresponds to the Pauli-Lubanski vector which is
boosted to the rest frame of the particle \cite{macfarlane_kinematics_1963},
\begin{align}
\left( \SNW \right)^j = \frac{1}{m} \left(L^{-1}_p W\right)^j,
\end{align}
where $L^{-1}_p$ is the boost that takes momentum $p$ to the rest system of
the particle, $L^{-1}_p p = (m, 0, 0, 0)$. We will use the Newton-Wigner spin
observable $\SNW$ throughout the paper to characterize the spin of the
particles.

Since we are working in the Wigner representation, we need to express
$\SNW$ in that representation. The canonical form of the
infinitesimal generators $P^{\mu}$, $\mbf{M}$ and $\mbf{N}$ is as
follows \cite{macfarlane_kinematics_1963},
\begin{align}
P^\mu &= p^\mu, \nonumber\\
\mbf{M} &= -i \ppp \times \partial_{\ppp} + \mbf{S},\label
{eq:CanonicalGenerators} \\
\mbf{N} &= -i p^0 \partial_{\ppp} - \frac{\ppp \times \mbf{S}}{m + p^0},
\nonumber
\end{align}
where $\SSS = \tfrac{1}{2} \sssigma$ and $\sssigma = (\sigma_x, \sigma_y,
\sigma_z)$ are the Pauli matrices. Substituting the generators
(\ref{eq:CanonicalGenerators}) into (\ref{eq:PauliLubanskiVector}) and
(\ref{eq:NewtonWignerS}), we obtain for the Newton-Wigner observable
\begin{align}
\SNW = \frac{1}{2} \sssigma,
\end{align}
meaning that in the Wigner representation $\SNW$ is given by the standard Pauli
matrices.

\section{Spin entanglement}

To find how the entanglement of the spin degree of freedom changes in various
boost scenarios, we will calculate the boosted spin state $\rho^\Lambda_S$. The
two particle spin state can be written in the operator basis
\begin{align}
\rho_S^\Lambda = \frac{1}{4} \left( \id \otimes \id + \mathbf{r} \boldsymbol\sigma
\otimes \id + \id \otimes \mathbf{s} \boldsymbol\sigma + \sum_{i,j} t_{ij}
\sigma_i \otimes \sigma_j \right),
\end{align}
where the coefficients $\mathbf{r} = (r_x, r_y, r_z)$, $\mathbf{s} = (s_x, s_y,
s_z)$ and $t_{ij}$, $i, j \in \{ x, y, z \}$ are the expectation values of the
spin observables $\sssigma \otimes \idd$, $\idd \otimes \sssigma$ and
$\sigma_i \otimes \sigma_j$.
Since the total state of two particles includes momentum as well, i.e.\ it
lives in the space
\begin{align}
\hilb^1_p \otimes \hilb^1_\lambda \otimes \hilb^2_p \otimes \hilb^2_\lambda,
\end{align}
the expectation values of observables have the form
\begin{align}
\langle \mathbf{r} \rangle =  \MTR \left( \rho^\Lambda \, \id_p^1 \otimes \SNW^1 \otimes \id_p^2 \otimes \id_{\sigma}^2 \right), \nonumber \\
\langle \mathbf{s} \rangle =  \MTR \left( \rho^\Lambda \, \id_p^1 \otimes \id^1_{\sigma} \otimes \id_p^2 \otimes \SNW^2 \right), \\
\langle t_{ij} \rangle = \MTR \left( \rho^\Lambda \, \id_p^1 \otimes \SNW^1 \otimes \id_p^2 \otimes \SNW^2 \right), \nonumber
\end{align}
where the superscripts denote the first and the second particle, respectively,
and $\rho^\Lambda = \pouter{\Psi^\Lambda}{\Psi^\Lambda}$.

Entanglement will be quantified by using the concurrence $C(\rho)$. This is
necessary since the final spin state $\rho^\Lambda$ is generally mixed.
Concurrence of a bipartite state $\rho$ of two qubits is defined as
\begin{align}
C(\rho) = \mathrm{max} \{0, \lambda_1 - \lambda_2 -\lambda_3 - \lambda_4 \},
\end{align}
where the $\lambda_i$ are square roots of eigenvalues of a non-Hermitian matrix
$\rho \widetilde{\rho}$ in decreasing order and
\begin{align}
\widetilde{\rho} = \left( \sigma_y \otimes \sigma_y \right) \rho^* \left(
\sigma_y \otimes \sigma_y \right),
\end{align}
with $\sigma_y$ a Pauli matrix, is the spin-flipped state with the complex
conjugate $^*$ taken in the standard basis~\cite{wootters_entanglement_1998}.

\section{Thomas--Wigner rotation}\label{sec:ThomasWignerRotation}

\begin{figure*}[htb]
\centering
\includegraphics[width=0.9\textwidth]{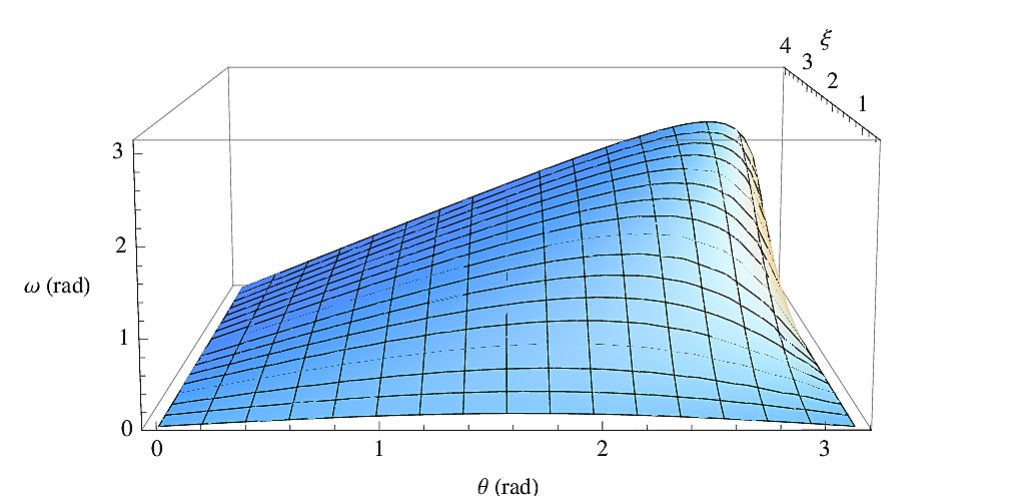}
\caption[TWR as a function of rapidity and boost angle.]%
{(Color online.) TWR $\omega$ as a function of rapidity $\xi$ and
boost angle $\theta$.}
\label{fig:figureTWRFunctionOfRapidityBoostAngle}
\end{figure*}

The TWR arises from the fact that the subset of Lorentz boosts does not form a
subgroup of the Lorentz group. Consider three inertial observers $O$, $O'$ and
$O''$ where $O'$ has velocity $\vvv_1$ relative to $O$ and $O''$ has $\vvv_2$
relative to $O'$. Then the combination of two canonical boosts
$\Lambda(\vvv_1)$ and $\Lambda(\vvv_2)$ that relates $O$ to $O''$ is in general
a boost \emph{and} a rotation,
\begin{align}
\Lambda(\vvv_2) \Lambda(\vvv_1) = R(\omega) \Lambda(\vvv_3),
\end{align}
where $R(\omega)$ is the TWR with angle $\omega$. To an observer $O$, the frame
of $O''$ appears to be rotated by $\omega$. We will immediately specialize to
massive systems, then $R(\omega) \in \mathrm{SO(3)}$ and $\omega$ is given by
\citep{rhodes_relativistic_2004,halpern_special_1968},
\begin{align}\label{eq:TWRHalfTanFormula}
\tan \frac{\omega}{2} = \frac{\sin \theta}{\cos \theta + D},
\end{align}
where $\theta$ is the angle between two boosts or, equivalently,
$\vvv_1$ and $\vvv_2$, and
\begin{align}\label{eq:theDFactor}
D = \sqrt{\left( \frac{\gamma_1 + 1}{\gamma_1 - 1} \right) \left( \frac{\gamma_2
+ 1}{\gamma_2 - 1} \right)},
\end{align}
with
$\gamma_{1,2} = (1 - v_{1,2}^2)^{-1/2}$ and $v_{1,2} = |\vvv_{1,2}|$.
We assume natural units throughout, $\hbar = c = 1$. The axis of rotation
specified by $\nnn = \vvv_2 \times \vvv_1 / |\vvv_2 \times \vvv_1|$ is
orthogonal to the plane defined by $\vvv_1$ and $\vvv_2$. Using rapidity
$\xi_{1, 2} = \mathrm{arctanh}\, \left| \vvv_{1, 2} \right|$ to represent the
magnitude of the boost and subsuming both under a single parameter $\xi = \xi_1
= \xi_2$, we show the dependence of the TWR on the boost angle $\theta$ and
$\xi$ in Fig.~\ref{fig:figureTWRFunctionOfRapidityBoostAngle}.
Two interesting characteristics are immediately noticeable. First, for any two
boosts at a fixed angle $\theta$, the TWR angle $\omega$ increases with $\xi$,
approaching a maximum value as boosts approach the speed of light. Second, the
angle $\theta$ at which the maximum TWR occurs depends on the magnitude of
$\xi$. It is worth noting that $\omega$ approaches the maximum value
$180\degree$ when boosts are almost opposite and approach the speed of light.
At lower boost magnitudes, maximum rotation occurs earlier.

\section{The model}

In this section, we will give a broad characterization of the models
to be studied below. More detailed discussion of the momentum and spin states
will be given in the next two sections.

We will assume throughout that initially the spin and momentum degrees of
freedom factorize,
\begin{align}\label{eq:initialGenericContinuousStateSpinMomentumFactorize}
\ket{\Psi} = \int \tum{p, q} \psi(\ppp, \qqq) \ket{\ppp, \qqq} \otimes
\ket{S},
\end{align}
where $\ket{S}$ is the spin state and momenta are taken to be combinations
of Gaussian wave packets in product forms. We start by considering product
momenta of the simplest form
\begin{align}
f_{\mathrm{PQ}}(\ppp, \qqq, \ppp_0, \qqq_0) =
  \left[ N(\sigma) \right]^{-\frac{1}{2}}
    g\!\left( \ppp, \ppp_0 \right)
    g\!\left( \qqq, \qqq_0 \right),
\end{align}
where $N(\sigma)$ is the normalization and $g(\ppp, \ppp_0)$ a Gaussian of width
$\sigma$ centered at $\ppp_0 = (p_{x0}, p_{y0}, p_{z0})$,
\begin{align}
g(\ppp, \ppp_0) &=
\left[
	\exp \left( -\frac{\left( p_x - p_{x0} \right)^2}{2 \sigma^2} \right)
	\exp \left( -\frac{\left( p_y - p_{y0} \right)^2}{2 \sigma^2} \right)
\right.
\nonumber\\
&\phantom{AA}
\left.
	\times\exp \left( -\frac{\left( p_z - p_{z0} \right)^2}{2 \sigma^2}
\right)
\right]^{\frac{1}{2}}.
\end{align}
There are two slightly different implementations of $f_{\mathrm{PQ}}$ that have
been discussed on a number of occasions.
When $\qqq_0 = -\ppp_0$ we get the familiar EPR--Bohm situation
\cite{einstein_can_1935, bohm_quantum_1951} with two particles moving in
opposite directions,
\begin{align}
f_{\mathrm{EPRB}}(\ppp, \qqq, \ppp_0, \qqq_0) = f_{\mathrm{PQ}}(\ppp, \qqq,
\ppp_0, -\ppp_0).
\end{align}
The other realization is described by
\begin{align}
f_C(\ppp, \qqq, \ppp_0, \qqq_0) =
f_{\mathrm{PQ}}(\ppp, \qqq, \ppp_{Z0}, \qqq_{Z0}),
\end{align}
which we will call axis centered momenta to signify that
the centers $\ppp_{Z0}$, $\qqq_{Z0}$ of Gaussians lie on a coordinate axis. With
no restriction of generality we take the coordinate axis to be the $z$-axis.

Further forms are motivated by observations we made at studying a single
particle system, namely, that superposed momenta give rise to maximal
entanglement between spin and momentum degrees of freedom. This suggests similar
momenta for two particles may give rise to interesting spin--spin effects as
well. We will consider a case involving two terms per particle,
\begin{align}\label{eq:GaussianFSigma}
f_{\Sigma}(\ppp, \qqq, \ppp_0, \qqq_0) &= \left[ N(\sigma)
\right]^{-\frac{1}{2}}
\left[ g\!\left( \ppp, \ppp_0 \right) + g\!\left( \ppp, -\ppp_0 \right) \right]
\nonumber\\
&\phantom{A}\times\left[ g\!\left( \qqq, \qqq_0 \right) + g\!\left( \qqq,
-\qqq_0 \right) \right]
\end{align}
and a more elaborate one with four terms,
\begin{align}
f_{\times}(\ppp, \qqq, \ppp_0, \qqq_0)
&= \left[ N(\sigma) \right]^{-\frac{1}{2}}
	\left[
		g(\ppp, \ppp_0) + g(\ppp, -\ppp_0)
	\right.
\nonumber\\
&\phantom{A}
	\left.
		+ g(\ppp, \ppp^{\perp}_{0}) + g(\ppp, -\ppp^{\perp}_0)
	\right]
	\left[g(\qqq, \qqq_0)
	\right.
\nonumber\\
&\phantom{A}
	\left.
		+ g(\qqq, -\qqq_0) + g(\qqq, \qqq^{\perp}_{0}) + g(\qqq,
-\qqq^{\perp}_0)
	\right],
\end{align}
where $\ppp^{\perp}_{0}$ is a momentum vector of the same magnitude but
orthogonal direction to $\ppp_{0}$ and similarly for $\qqq^{\perp}_{0}$ and
$\qqq_{0}$.

Boosts are always assumed to be in the $z$-direction, $\Lambda \equiv
\Lambda_z(\xi)$,
\begin{align}\label{eq:LorentzBoostPositiveZDirectionTwoParticles}
\Lambda =
\bpm
\cosh \xi   & \;\; &  0  & \;\; &  0  & \;\; &  \sinh \xi \;\; \\
0   & \;\; &  1  & \;\; &  0  & \;\; &  0 \;\; \\
0   & \;\; &  0  & \;\; &  1  & \;\; &  0 \;\; \\
\sinh \xi   & \;\; &  0  & \;\; &  0  & \;\; &  \cosh \xi \;\;
\epm .
\end{align}
This implies that the unitary representation of the TWR
acting on the one particle subsystem takes the form \cite{halpern_special_1968}
\begin{align}\label{eq:HalpernSpinHalfBoostZDirectionTwoParticles}
D[W(\Lambda, \ppp)] =
\bpm
\alpha  &   \beta (p_x - \II p_y) \\
-\beta (p_x + \II p_y)  &   \alpha
\epm,
\end{align}
where we have denoted
\begin{align}
\alpha &= \sqrt{\frac{E + m}{E^{\Lambda} + m}} \left(  \cosh \frac{\xi}{2} +
\frac{p_z}{E + m} \sinh \frac{\xi}{2} \right),
\nonumber \\
{\footnotesize\phantom{a}} \nonumber \\
\beta &= \frac{1}{\sqrt{(E + m) (E^{\Lambda} + m)}} \sinh \frac{\xi}{2} ,
\end{align}
with $\xi$ being the rapidity of the boost in the $z$-direction,
and
\begin{align}
E^{\Lambda} = E \cosh \xi + p_z \sinh \xi.
\end{align}
Because the expression of the boosted spin state in is too complex to be
tackled by analytic methods, we will resort to numerical treatment in
determining the concurrence and the orbits of states. No numerical
approximations are involved except for the discretization of the momentum
space.

\section{Momenta and spin rotations}\label{sec:momentaAndSpinRotations}

Although we have now specified the generic forms that momenta will take, the
particular geometry they might realize is still undetermined. For instance, the
geometric momenta $\ppp_0$ and $\qqq_0$ that specify the centers of Gaussians in
$f_{\Sigma}$, Eq.~(\ref{eq:GaussianFSigma}), may lie along the same momentum
axis, or they may lie along orthogonal axes. They will, correspondingly,
generate different types of rotations on the spins. In this section, we will
focus on how the generic Gaussian states can be implemented by particular
momenta and relate them to different types of rotations generated on spins. To
make the discussion perspicuous, we will use discrete momentum states, denoted
by $\ket{M}$, that have the same form and subscripts as the continuous ones.

\begin{figure*}
	\centering
	\begin{subfigure}[t]{0.28\textwidth}
		\centering
		\includegraphics[width=\textwidth]{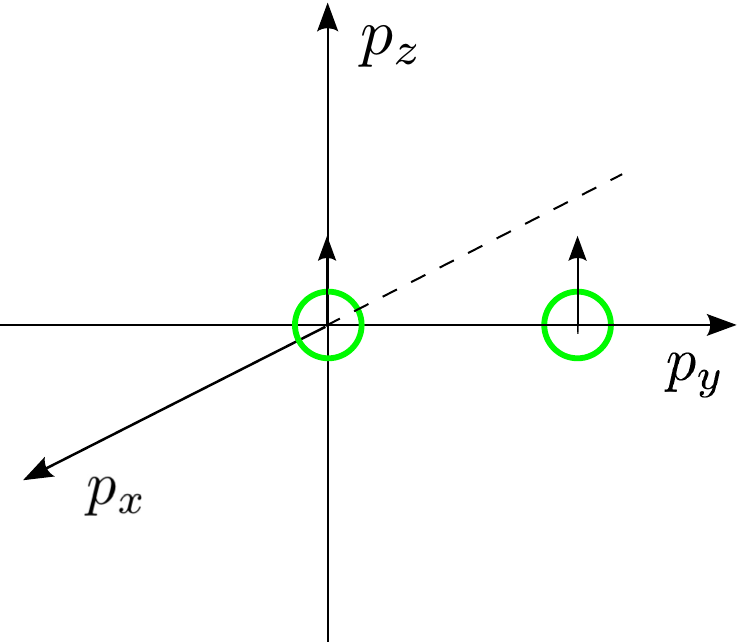}
		\caption{}
		\label{fig:X4a}
	\end{subfigure}
	\hspace{2em}
	\begin{subfigure}[t]{0.28\textwidth}
		\centering
		\includegraphics[width=\textwidth]{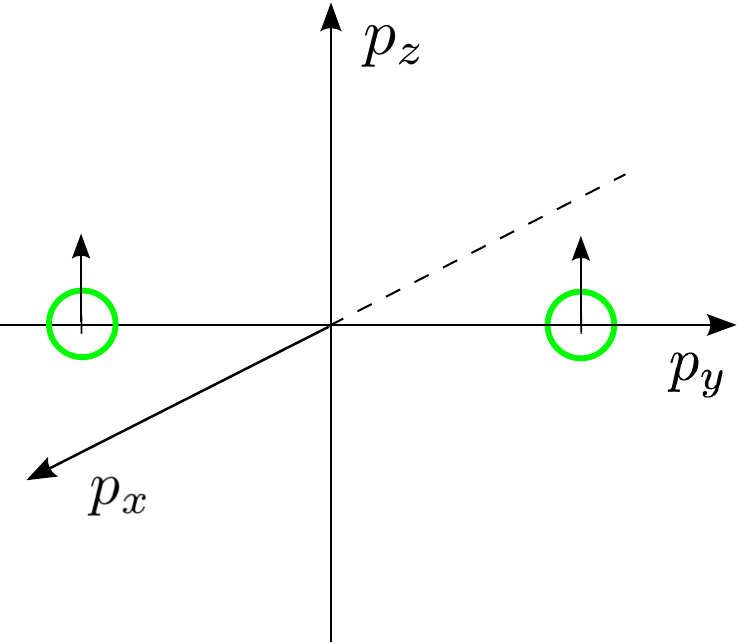}
		\caption{}
		\label{fig:X4b}
	\end{subfigure}
	\hspace{2em}
	\begin{subfigure}[t]{0.28\textwidth}
		\centering
		\includegraphics[width=\textwidth]{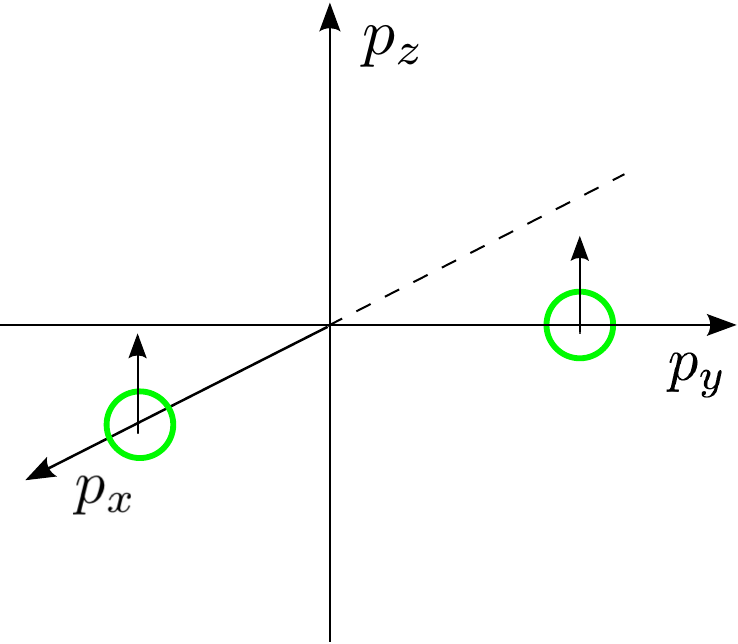}
		\caption{}
		\label{fig:X4c}
	\end{subfigure}
\caption{(Color online.) Schematic illustration. Examples of
geometric configurations of Gaussian momenta (green circles) for
realizations of different types of rotations on spins, with (a) $R_i
\otimes \idd$, (b) $R_i \otimes R_i$, (c) $R_i \otimes R_j, i \ne j$.
The $z$-projection of the spin field is indicated by an arrow at the
Gaussian.}
	\label{fig:X4}
\end{figure*}

Momenta of both particles may be aligned along the same axes, for instance two
particles can be in a superposition of momenta along the $x$-axis, yielding the
state,
\begin{align}\label{eq:momentaAlongTheSameAxis}
\ket{M_\Sigma^{XX}} = \frac{1}{2} \left( \ket{\ppp_x} + \ket{-\ppp_x} \right)
\left( \ket{\qqq_x} + \ket{-\qqq_x} \right).
\end{align}
Or momenta of both particles may be aligned along different axes, for instance
the first particle might be in a superposition of momenta along the $x$-axis and
the second particle in a superposition along the $y$-axis,
\begin{align}\label{eq:momentaAlongDifferentAxis}
\ket{M_\Sigma^{XY}} = \frac{1}{2} \left( \ket{\ppp_x} + \ket{-\ppp_x} \right)
\left( \ket{\qqq_y} + \ket{-\qqq_y} \right).
\end{align}
Following the assumption
(\ref{eq:initialGenericContinuousStateSpinMomentumFactorize}) above that
initially spin and momentum factorize,
\begin{align}
\ket{\Psi} = \ket{M} \otimes \ket{S},
\end{align}
and substituting momentum $\ket{M_\Sigma^{XX}}$ into
(\ref{eq:unitaryActionOnFiberTwoParticles})
we obtain the boosted state
\begin{align}\label{eq:momentumExplicitlyWithURotations}
\ket{\Psi^{\Lambda}} = &\frac{1}{2} \biggl\{
	\ket{\Lambda_z \ppp_x, \Lambda_z \qqq_x} \, D\!\left[W(\Lambda_z,
\ppp_x)\right] \otimes D\!\left[W(\Lambda_z, \qqq_x)\right] \biggr. \nonumber\\
	&+ \ket{\Lambda_z \ppp_x, -\Lambda_z \qqq_x} \, D\!\left[W(\Lambda_z,
\ppp_x)\right] \otimes D\!\left[W(\Lambda_z, -\qqq_x)\right] \nonumber\\
	&+ \ket{-\Lambda_z \ppp_x, \Lambda_z \qqq_x} \, D\!\left[W(\Lambda_z,
-\ppp_x)\right] \otimes D\!\left[W(\Lambda_z, \qqq_x)\right] \nonumber\\
	&+ \biggl.  \ket{-\Lambda_z \ppp_x, -\Lambda_z \qqq_x} \,
D\!\left[W(\Lambda_z, -\ppp_x)\right] \nonumber\\
	&\otimes D\!\left[W(\Lambda_z, -\qqq_x)\right] \biggr\}
	\ket{S},
\end{align}
where for the sake of concreteness we have taken the boost to be in the
$z$-direction. Now the operators $D\!\left[W(\Lambda, \ppp)\right]$ for the
unitary representation of the Wigner rotation in this expression are given in
terms of the momenta, the direction of boost and rapidity, that is, variables
which specify the configuration of the boost in the physical three space.
Formally they are $\mathrm{SU}(2)$ operators parameterized by the latter three
quantities. However, as long as our main interest lies in clarifying what kind
of rotations boosts induce on spins we can simplify the notation and write
$R_Y(\omega)$ instead of $D[W(\Lambda_z, \ppp_x)]$, meaning that the spin is
rotated around the $y$-axis by angle $\omega$. Using this,
Eq.~(\ref{eq:momentumExplicitlyWithURotations}) can be written as
\begin{align}\label{eq:momentumExplicitlyWithRRotations}
\ket{\Psi^{\Lambda}} = &\frac{1}{2} \bigl[
	\ket{\Lambda_z \ppp_x, \Lambda_z \qqq_x} \, R_Y(\omega) \otimes
R_Y(\chi) \big. \nonumber\\
	&+ \ket{\Lambda_z \ppp_x, -\Lambda_z \qqq_x} \, R_Y(\omega) \otimes
R_Y(-\chi) \nonumber\\
	&+ \ket{-\Lambda_z \ppp_x, \Lambda_z \qqq_x} \, R_Y(-\omega) \otimes
R_Y(\chi) \nonumber\\
	&+ \big.  \ket{-\Lambda_z \ppp_x, -\Lambda_z \qqq_x} \, R_Y(-\omega)
\otimes R_Y(-\chi)
	\bigr] \ket{S}.\nonumber\\
	&\phantom{A}
\end{align}
Thus we see that the momenta $\ket{M_\Sigma^{XX}}$ generate rotations of the
form
\begin{align}
R_Y(\pm\omega) \otimes R_Y(\pm\chi), \quad R_Y(\pm\omega) \otimes R_Y(\mp\chi)
\end{align}
on the spin state. In the same vein, if the momenta are given by
$\ket{M_\Sigma^{XY}}$ the $z$-boosted state will have terms that generate
rotations
\begin{align}
R_Y(\pm\omega) \otimes R_X(\pm\chi), \quad R_Y(\pm\omega) \otimes R_X(\mp\chi)
\end{align}
on the spin state. Following considerations along these lines we see that by
taking momenta along different combinations of axes for product momenta, one
obtains three different types of rotations that can occur on the spin state,
\begin{align}\label{eq:rotationTypesGeneral}
\mathrm{(i) }\;\;  &R_i \otimes \idd,  \nonumber\\
\mathrm{(ii) }\;\; &R_i \otimes R_i,  \\
\mathrm{(iii) }\;\; &R_i \otimes R_j, \quad i \ne j, \nonumber
\end{align}
where $i, j \in \{ X, Y, Z \}$ and each type of rotation can be realized by some
set of suitably chosen momenta, see Fig.~\ref{fig:X4}.
For instance, we saw that $R_i \otimes R_i$ is instantiated by $R_Y \otimes R_Y$
when the momenta are given by the product state $\ket{M_\Sigma^{XX}}$ and the
boost is in the $z$-direction. Another implementation of the same type is $R_X
\otimes R_X$ when the momenta are again a product but located along the
$y$-axis, $\ket{M_\Sigma^{YY}}$, and the boost is in the $z$-direction.

We will next give a few examples of momenta and boost geometries that implement
the different types of rotations listed in~(\ref{eq:rotationTypesGeneral}).

\paragraph{Type $R_i \otimes \idd$.} In this scenario, only the first particle
undergoes rotation. The momentum of the second particle is chosen so that it
leaves the spin alone. Denoting such a momentum by $\ket{0}$, the following
pairs of boosts and momenta listed on the left hand side generate rotations
given on the right hand side,
\begin{align}\label{eq:realizationOfR_i_1_}
\Lambda_z \, , \ket{\ppp_y, 0}  \longmapsto  R_X \otimes \idd, \nonumber\\
\Lambda_z \, , \ket{\ppp_x, 0}  \longmapsto  R_Y \otimes \idd, \\
\Lambda_y \, , \ket{\ppp_x, 0}  \longmapsto  R_Z \otimes \idd. \nonumber
\end{align}

\paragraph{Type $R_i \otimes R_i$.} For scenarios in which both particles are
rotated around the same axis but not necessarily in the same direction, we
obtain the following boosts and momenta,
\begin{align}\label{eq:realizationOfR_i_R_i_}
\Lambda_z \, , \ket{\ppp_y, \qqq_y}   \longmapsto   R_X \otimes R_X, \nonumber\\
\Lambda_z \, , \ket{\ppp_x, \qqq_x}   \longmapsto   R_Y \otimes R_Y, \\
\Lambda_y \, , \ket{\ppp_x, \qqq_x}   \longmapsto   R_Z \otimes R_Z. \nonumber
\end{align}

\paragraph{Type $R_i \otimes R_j$, $i \ne j$.} Scenarios where particles undergo
rotations around different axes can be realized by
\begin{align}\label{eq:realizationOfR_i_R_j_}
\Lambda_y \, , \ket{\ppp_z, \qqq_x}   \longmapsto   R_X \otimes R_Z, \nonumber\\
\Lambda_z \, , \ket{\ppp_y, \qqq_x}   \longmapsto   R_X \otimes R_Y, \\
\Lambda_x \, , \ket{\ppp_z, \qqq_y}   \longmapsto   R_Y \otimes R_Z. \nonumber
\end{align}

\vspace{1em}
These scenarios admit an obvious generalization. By choosing momenta and boosts
appropriately, one can consider single particle rotations around an arbitrary
axis $\nnn = (n_x, n_y, n_z)$. This leads to combinations of generic rotations
$R_{\nnn_1} \otimes R_{\nnn_2}$ for two particle systems, opening up a wide
avenue of research. However, when surveying the landscape for the first time,
we would like to keep the situation tractable by confining attention to the
cases listed above and leave a more general approach for another occasion.

\section{Spin state and its visualization}
\label{sec:spinStateAndItsVisualization}

We will next characterize the spin state of the system. Most previous work has
focussed on the Bell states,
\begin{align}
\ket{\Phi_{\pm}} = \frac{1}{\sqrt{2}} \left( \ket{00} \pm \ket{11}\right), \quad
\ket{\Psi_{\pm}} = \frac{1}{\sqrt{2}} \left( \ket{01} \pm \ket{10}\right),
\end{align}
the maximally entangled bipartite states of two level systems. Understanding
their behavior in relativity is important for quantum information and we will
follow suit in this paper
\footnote{From a more general perspective, it is interesting to consider mixed
states as well. See \cite{palge_behavior_2015} for an extension to the Werner
states.}.
As regards the geometric configuration, we will assume throughout that the spins
are aligned with the $z$-axis irrespective of the direction of the boost.
We adopt the convention that $\ket{0}$ signifies the `up' spin and $\ket{1}$ the
`down' spin.

In order to gain a better understanding of the state change of a single qubit,
one commonly uses visualization in terms of the Bloch sphere. Visualization of
two qubits, however, is in general impossible since one needs $15$ real
parameters to characterize the density matrix. However, some cases still allow
for a representation in three space, for instance when the state is restricted
to evolve in a subspace of few dimensions. Fortunately this turns out to be the
case for our system.

It is useful to work in the Hilbert-Schmidt space of operators $B(\hilb)$,
defined on the Hilbert space $\hilb$ with \mbox{$\text{dim} = N$}
\cite{bengtsson_geometry_2006}. $B(\hilb)$ becomes a Hilbert space of $N^2$
complex dimensions when equipped with a scalar product defined as
\mbox{$\langle A | B \rangle = \mtr(A^{\dagger}B)$}, with $A, B \in B(\hilb)$,
where the squared norm is \mbox{$\| A \|^2 = \mtr(A^{\dagger}A)$}. The vector
space of Hermitian operators is an $N^2$ real-dimensional subspace of
Hilbert-Schmidt space which can be coordinatized using a basis that consists of
identity operator and the generators of $\sun$. For a qubit $N = 2$ and we
obtain the familiar Bloch ball. For a bipartite qubit system $N = 4$, $B(\hilb)
= B(\hilb_A) \otimes B(\hilb_B)$ where $\hilb_i$ is the single particle space,
and we can use a basis whose elements are the tensor products $\{ \idd \otimes
\idd, \idd \otimes \boldsymbol\sigma, \boldsymbol\sigma \otimes \idd,
\boldsymbol\sigma \otimes \boldsymbol\sigma \}$, where $\boldsymbol\sigma =
\left( \sigma_x, \sigma_y, \sigma_z \right)$ is the vector of Pauli operators.
The density operator for a $2 \times 2$ dimensional system can be written in
the general form,
\begin{align}
\rho = \frac{1}{4} \left( \idd \otimes \idd + \mathbf{r} \boldsymbol\sigma
\otimes \idd + \idd \otimes \mathbf{s} \boldsymbol\sigma + \sum_{i,j} t_{ij}
\sigma_i \otimes \sigma_j \right),
\end{align}
where the coefficients $\mathbf{r} = (r_x, r_y, r_z)$, $\mathbf{s} = (s_x, s_y,
s_z)$ and $t_{ij}$, $i, j \in \{ x, y, z \}$ are the expectation values of the
operators $\boldsymbol\sigma \otimes \idd$, $\idd \otimes
\boldsymbol\sigma$ and $\sigma_i \otimes \sigma_j$.

\begin{figure}[htb]
\centering
\includegraphics[width=0.4\textwidth]{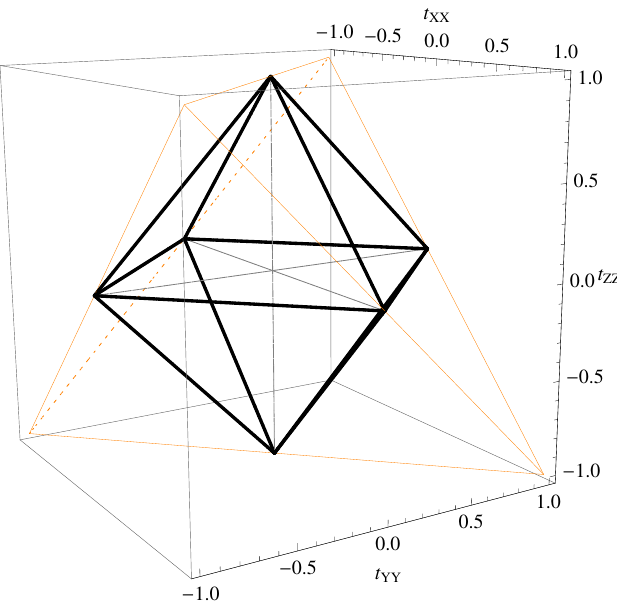}
\caption[The geometry of Bell diagonal states.]{(Color online.) The
geometry of Bell diagonal states. The vertices of the tetrahedron
$\mathcal{T}$ (thin yellow lines) correspond to the four Bell states
$\ket{\Phi_+}$, $\ket{\Phi_-}$, $\ket{\Psi_+}$, and $\ket{\Psi_-}$.
Convex combinations of projectors on the Bell states, the Bell
diagonal states, lie on or in the tetrahedron (thick black lines). A
Bell diagonal state is separable iff it lies in the double pyramid
formed by the intersection of the tetrahedron $\mathcal{T}$ and its
reflection through the origin~$\mathcal{-T}$.}
\label{fig:bipartiteTetrahedronGeneral}
\end{figure}

For the projectors on the Bell states $s_i = r_i = 0$ and the matrix $t_{ij}$ is
diagonal. This implies we only need to consider the values of diagonal
components $t_{ii}$ which constitute a vector in $3$-dimensional space, allowing
us to represent the states in Euclidean three
space~\cite{bertlmann_geometric_2002}. The Bell states correspond to vectors,
\begin{align}
t_{\Phi_+} = \left( 1, -1, \phantom{-}1 \right), \quad\quad
t_{\Phi_-} = \left( -1, \phantom{-}1, \phantom{-}1 \right), \nonumber \\
t_{\Psi_+} = \left( 1, \phantom{-}1, -1 \right), \quad\quad
t_{\Psi_-} = \left( -1, -1, -1 \right),
\end{align}
which, in turn, correspond to the vertices of a tetrahedron $\mathcal{T}$ in
Fig.~\ref{fig:bipartiteTetrahedronGeneral}. By taking convex combinations of
these, one obtains further diagonal states; the set of all such states is called
\emph{Bell-diagonal} and is represented by the (yellow) tetrahedron
$\mathcal{T}$ in Fig.~\ref{fig:bipartiteTetrahedronGeneral}.
The set of separable states forms a double pyramid, an octahedron, in the
tetrahedron. The octahedron is given by the intersection of $\mathcal{T}$ with
its reflection through the origin, $-\mathcal{T}$. The maximally mixed state
$\frac{1}{4} \idd_4$ has coordinates $(0, 0, 0)$ and it lies at the origin. The
entangled states are located outside the octahedron in the cones of the
tetrahedron, see Fig.~\ref{fig:bipartiteTetrahedronGeneral}.

We can now visualize the behavior of spin by calculating the coefficients
$t_{ii}$ under a given rotation as a function of rapidity $\xi$,
\begin{align}\label{eq:stateAsThreeVector}
t(\xi) = \left(t_{xx}, t_{yy}, t_{zz}\right),
\end{align}
where
\begin{align}\label{eq:coefficientsOfStateVector}
t_{ii} = \MTR\!\left[\rho^{\Lambda}_S(\xi)\, \sigma_i \otimes
\sigma_i\right],\quad i \in \{ x, y, z \},
\end{align}
and $\rho^{\Lambda}_S(\xi)$ is the boosted spin state. The resulting set of
three vectors
\begin{align}
\Gamma\!\left[\rho^{\Lambda}_S(\xi)\right] = \{ t(\xi) \;|\; \xi \in [0,
\xi_{\text{max}}] \}
\end{align}
we call an \emph{orbit} of a given initial state. It can be represented as a
curve in three space in the manner described above.

\section{Product momenta $f_{\mathrm{EPRB}}$}\label{sec:EPRProductMomenta}

\begin{figure}[htb]
\centering
\includegraphics[width=0.5\textwidth]{%
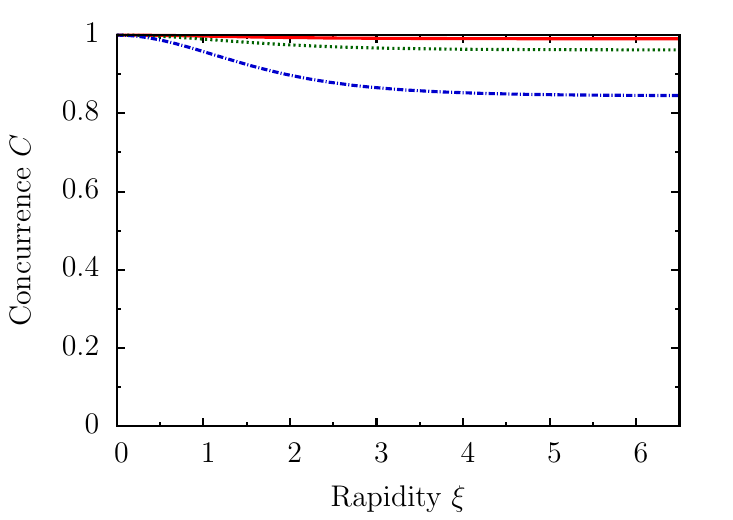}
\caption%
[Spin concurrence for Gaussian momenta $f_{\mathrm{EPRB}}$ with $\sigma / m = 1,
2, 4$.]
{(Color online.) Spin concurrence for Gaussian momenta
$f_{\mathrm{EPRB}}$ with $\sigma / m = 1, 2, 4$ and centers of
Gaussians lying on the $x$-axis at $\pm \ppp_{X0} = (\pm 17.13, 0,
0)$. Data for $\sigma / m = 1$ is shown with the solid red line,
$\sigma / m = 2$ green dotted line and $\sigma / m = 4$ blue
dot-dashed line.
}%
\label{fig:concurrence_BELL_continuous_PRODUCT_EPR_p_q_sigma_1_2_4}
\end{figure}

We begin by considering product momenta of the simplest form
\begin{align}
f_{\mathrm{EPRB}}(\ppp, \qqq, \ppp_0, \qqq_0) =
  \left[ N(\sigma) \right]^{-\frac{1}{2}}
    g\!\left( \ppp, \ppp_0 \right)
    g\!\left( \qqq, -\ppp_0 \right),
\end{align}
which represent the EPR--Bohm scenario where two particles move in opposite
directions $\ppp_0$ and $-\ppp_0$. Early discussion focussed on momentum delta
states and concluded that spin entanglement of a Bell state was left invariant
Lorentz boosts \cite{alsing_lorentz_2002, terashima_relativistic_2003}. We
reproduce the case of delta momentum by approximating it with a narrow Gaussian
of width $\sigma / m = 1$. In order to study wavepackets of larger widths, we
also calculate $\sigma / m = 2, 4$. The concurrence for all three cases is shown
in Fig.~\ref{fig:concurrence_BELL_continuous_PRODUCT_EPR_p_q_sigma_1_2_4}.
Unfortunately, the spin orbit cannot be visualized since it is not Bell
diagonal.

The narrow momenta $\sigma / m = 1$ which approximate the delta state confirm
the results obtained by \cite{alsing_lorentz_2002, terashima_relativistic_2003},
namely, that boosts leave the entanglement of a maximally entangled Bell state
invariant. Larger widths, however, show a decrease of entanglement, which grows
with the width and magnitude of the boost.

To analyze the behavior, we will resort to the simple discrete model used above
when discussing the relation between momenta and rotations. We can approximate
the narrow momenta $\sigma / m = 1$ by a single momentum term and write the
total state of the boosted particle in discrete form as in
(\ref{eq:momentumExplicitlyWithURotations}),
\begin{align}
\ket{\Psi^{\Lambda}} &= \ket{\Lambda_z \ppp_0, -\Lambda_z \ppp_0} \,
D\!\left[W(\Lambda_z, \ppp_0)\right] \nonumber\\
&\phantom{AAAAAAAA}\otimes D\!\left[W(\Lambda_z, -\ppp_0)\right] \ket{\Phi_+}.
\end{align}
This shows that the boost generates a local unitary transform of the form $D_1
\otimes D_2$ on the spin state $\ket{\Phi_+}$. Since the degree of entanglement
of any Bell state is left invariant by such a transform, Lorentz boosts do not
change the entanglement between spins in this case.

Systems with widths $\sigma / m = 2, 4$, however, display loss of spin
entanglement. This is because they cannot be modeled using a single momentum
term. A larger width means that, in analogy with
Eq.~(\ref{eq:momentumExplicitlyWithURotations}), the discrete model now
consists of several momenta and the boosted state involves many rotation
operators acting on the spin state. Calculating the boosted spin state, we
obtain
\begin{align}
\rho_{\Phi+}^{\Lambda}
&= \sum_{\ppp, \qqq} |f_{\mathrm{EPRB}}(\ppp, \qqq)|^2
D\!\left[W(\Lambda, \ppp, \qqq)\right] \rho_{\Phi+} \,
D^{\dagger}\!\left[W(\Lambda, \ppp, \qqq)\right],
\end{align}
where $f_{\mathrm{EPRB}}$ is centered at $\ppp_0$ and $-\ppp_0$, respectively,
$\rho_{\Phi+} = \pouter{\Phi_+}{\Phi_+}$, and we have abbreviated
$D\!\left[W(\Lambda, \ppp, \qqq)\right] \equiv D\!\left[W(\Lambda, \ppp)\right]
\otimes D\!\left[W(\Lambda, \qqq)\right]$.
The final spin state $\rho_{\Phi+}^{\Lambda}$ is in general a mixed state whose
entanglement has changed as a result of the boost. Based on the discrete model,
one would expect that larger widths lead to bigger changes when rapidity
increases because, roughly, the rotations generated by different momenta diverge
more than in the case of narrow momenta. Indeed, the plots of $\sigma / m = 2,
4$ in Fig.~\ref{fig:concurrence_BELL_continuous_PRODUCT_EPR_p_q_sigma_1_2_4},
which have been obtained using numerical methods, confirm this intuition.

\section{Product momenta $f_{\Sigma}$}\label{sec:productMomenta}

\begin{figure}[htb]
\centering
\includegraphics[width=0.35\textwidth]{%
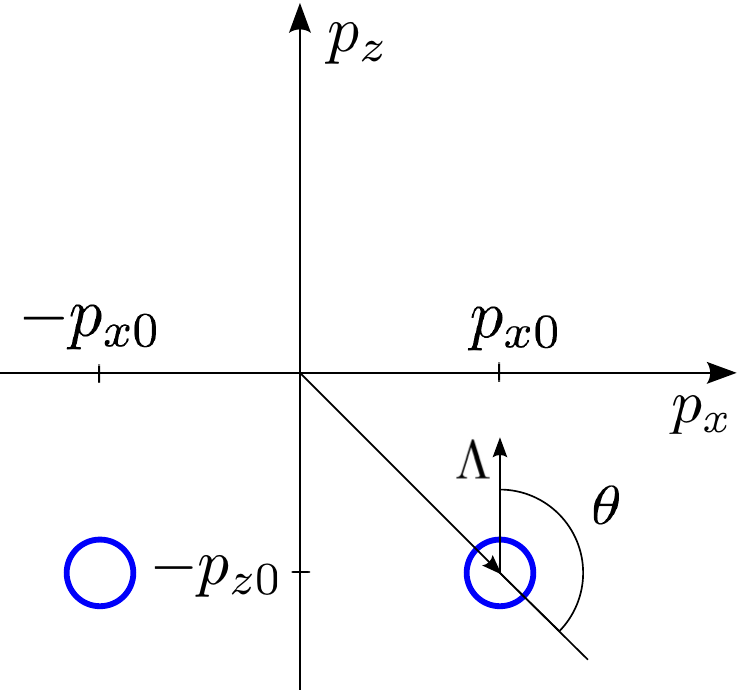}
\caption{(Color online.) Schematic illustration of a boost at a large
angle $\theta$. Gaussian momenta (shown as blue circles) are located
at $(\pm p_{x0}, 0, -p_{z0})$. Boost $\Lambda$ is in the positive
$z$-direction.}
\label{fig:X5}
\end{figure}

In this section we will focus on spin rotations generated by product momenta of
the form $f_{\Sigma}$. In order to study the maximum range of phenomena that
Lorentz boosts can exhibit we will choose boost scenarios with large boost
angles and momenta so that the spins undergo large TWR when boosts approach the
speed of light. To this end, we will assume that the centers of the Gaussians
are given by geometric vectors $\pm \ppp_{X0} = (\pm 17.13, 0, -98.5)$ and $\pm
\ppp_{Y0} = (0, \pm 17.13, -98.5)$, see Fig.~\ref{fig:X5}.
This corresponds to the maximum TWR of $163\degree$ at large boosts $\xi = 6.5$.

\subsection{Case $R_i \otimes \idd$}\label{sec:bellContinuousMomentaR_i_1}

It is not easy to implement rotations of type $R_i \otimes \idd$ in the
continuous regime as long as we are concerned with the physical situation where
the observer moves relative to both particles. The problem lies in realizing
the identity map. Even if we find a scenario where boosts leave alone a
momentum given by a delta state, the non-zero width of the wave packet
guarantees that this will not apply to the whole wave packet. Some parts of the
wave packet will necessarily induce non-trivial transformations on the spin
state as we learned in studying the continuous momentum models of a single
particle in \cite{palge_generation_2012}. We will thus adopt the strategy of
constructing a model that approximates the identity map to as high a degree as
possible by minimizing the effect of boost on the spin of the second particle.

Above we fixed the boost to be always in the $z$-direction. In order to realize
the \mbox{$R_i \otimes \idd$} rotations, we will take the momentum of the first
particle to lie in the
\plane{z}{x}
with \mbox{$\pm\ppp_0 = \pm \ppp_{X0}$}, while the momentum of the second
particle is located at the origin of the
\plane{x}{y}
with the $z$-component equal to that of the first particle, \mbox{$\qqq_0 = (0,
0, -98.5)$}. Since the momentum of the second particle is aligned with the
direction of the boost, the resulting rotation of the spin field approximates
the identity map.

\begin{figure*}
	\centering
	\begin{subfigure}[t]{0.7\textwidth}
		\centering

\includegraphics[width=\textwidth]{%
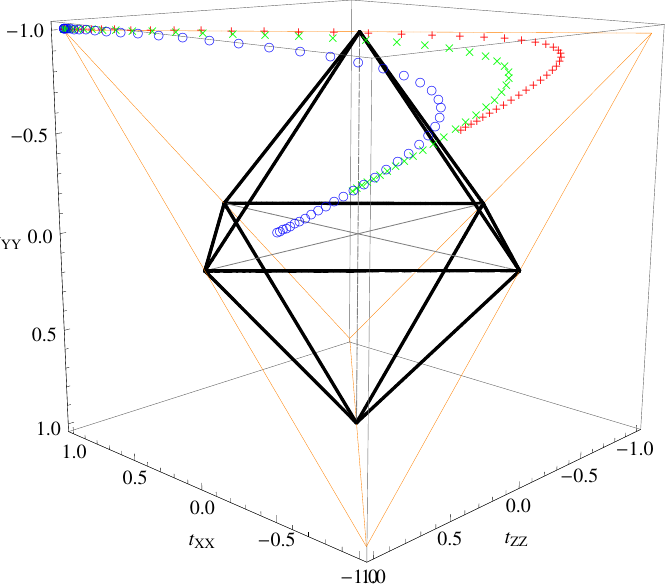}
		\caption{}

\label{fig:orbit_BELL_continuous_PRODUCT_momenta_R_i_1_sigma_1_2_4}
	\end{subfigure}
	\hspace{2em}
	\begin{subfigure}[t]{0.7\textwidth}
		\centering

\includegraphics[width=\textwidth]{%
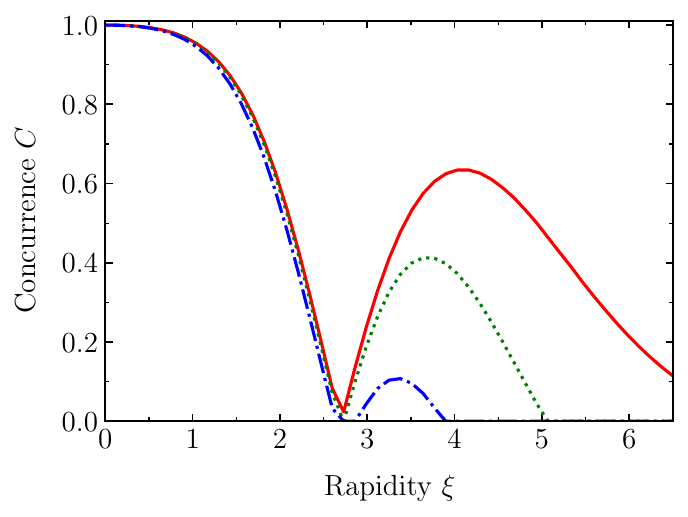}
		\caption{}
\label{fig:concurrence_BELL_continuous_PRODUCT_momenta_R_i_1_sigma_1_2_4}
	\end{subfigure}
\caption[(a) orbit and (b) concurrence under
$R_i \otimes \idd$ for Gaussian momenta $f_{\Sigma}$ with $\sigma / m
= 1, 2, 4$.]{(Color online.) Spin (a) orbit and (b) concurrence under $R_i \otimes
\idd$ for Gaussian momenta $f_{\Sigma}$ with $\sigma / m = 1, 2, 4$.
Data for
$\sigma / m = 1$ is shown with (a) the red ``+'' and (b) the red solid line,
$\sigma / m = 2$ with (a) the green ``$\times$'' and (b) the green dotted line,
$\sigma / m = 4$ with (a) the blue ``$\bigcirc$'' and (b) the blue dot-dashed line.
(a)~Initial state $\ket{\Phi_+}$ corresponds
to vertex $(1, -1, 1)$.}
	\label{fig:BELL_continuous_PRODUCT_momenta_R_i_1_sigma_1_2_4}
\end{figure*}

We plot the orbit of the spin state along with its concurrence in
Fig.~\ref{fig:BELL_continuous_PRODUCT_momenta_R_i_1_sigma_1_2_4}.
It is evident that visualization of the orbit provides valuable insight into the
behavior of the state, as well as explaining the behavior of entanglement. Let
us begin by considering the case $\sigma / m = 1$, shown red in
Fig.~\ref{fig:orbit_BELL_continuous_PRODUCT_momenta_R_i_1_sigma_1_2_4}.
Initially the state is at rest, represented by the state $\ket{\Phi_+}$ at the
vertex $(1, -1, 1)$. When boosts begin to increase, the state moves towards the
center of the face, reaching a separable state $(0, -1, 0)$ at about $\xi =
2.7$. Correspondingly, the concurrence initially takes value $1$, decreasing
monotonically with the increase of boosts. It vanishes at about $\xi = 2.7$ when
the state hits the separable region.

When boosts become larger than $2.7$, the spin of the first particle is rotated
even further, and the system becomes again entangled, with the orbit moving
towards the vertex $(-1, -1, -1)$ which represents the Bell state
$\ket{\Psi_-}$. However, the revival of entanglement stops short of reaching the
value $0.64$ for concurrence. Concurrence starts to decrease when $\xi$ becomes
larger than $4.16$.

While the states with $\sigma / m = 2$ and $\sigma / m = 4$ display similar
qualitative behavior, their orbits lie increasingly more in the region of
separable states as $\sigma / m$ becomes larger, see
Fig.~\ref{fig:orbit_BELL_continuous_PRODUCT_momenta_R_i_1_sigma_1_2_4}. As a
consequence, the revival of concurrence becomes less pronounced, recovering
only briefly for $\sigma / m = 4$ in the interval $\xi \in [2.9, 3.9]$ and
vanishing thereafter as the state enters the octahedron of separable states.

\subsection{Case $R_i \otimes R_i$}\label{sec:f-Sigma-R_i_times_R_i}

\begin{figure*}
	\centering
	\begin{subfigure}[t]{0.7\textwidth}
		\centering

\includegraphics[width=\textwidth]{%
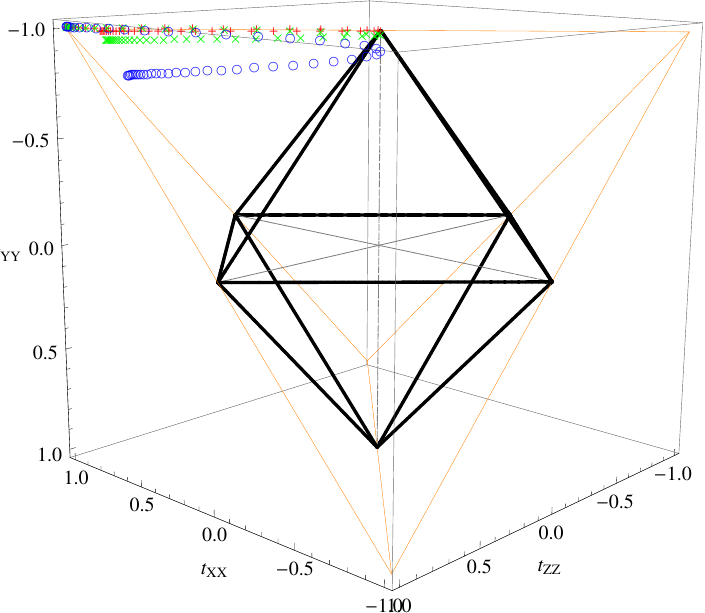}
		\caption{}

\label{fig:orbit_BELL_continuous_PRODUCT_momenta_R_i_R_i_sigma_1_2_4}
	\end{subfigure}
	\hspace{1em}
	\begin{subfigure}[t]{0.7\textwidth}
		\centering

\includegraphics[width=\textwidth]{%
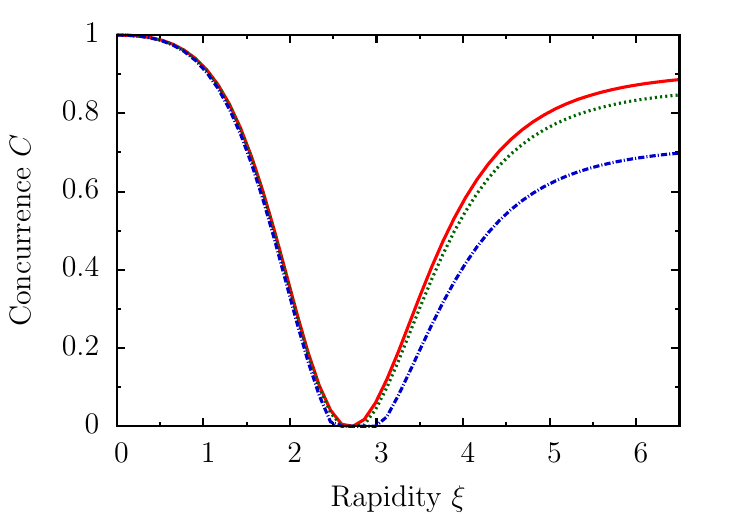}
		\caption{}

\label{fig:concurrence_BELL_continuous_PRODUCT_momenta_R_i_R_i_sigma_1_2_4}
	\end{subfigure}
	\caption%
[Spin (a) orbit and (b) concurrence under $R_i \otimes R_i$ for Gaussian momenta
$f_{\Sigma}$ with \mbox{$\sigma / m = 1, 2, 4$}.]%
{(Color online.) Spin (a) orbit and (b) concurrence under $R_i
\otimes R_i$ for Gaussian momenta with \mbox{$\sigma / m = 1, 2, 4$}.
Product momenta are given by $f_{\Sigma}$. Data for $\sigma / m = 1$
is shown with (a) the red ``+'' and (b) the red solid line, $\sigma /
m = 2$ with (a) the green ``$\times$'' and (b) the green dotted line,
$\sigma / m = 4$ with (a) the blue ``$\bigcirc$'' and (b) the blue
dot-dashed line. (a)~Initial state $\ket{\Phi_+}$ corresponds to
vertex $(1, -1, 1)$.}
\label{fig:BELL_continuous_PRODUCT_momenta_R_i_R_i_sigma_1_2_4}
\end{figure*}

To implement the type of rotation where both particles undergo rotation around
the same axis, the momenta $\ppp_0$ and $\qqq_0$ need to lie in the same boost
plane. Since the boost is in the $z$-direction, we will assume that the
Gaussians are centered at the geometric vectors \mbox{$\pm\qqq_0 = \pm\ppp_0 =
\pm\ppp_{X0}$}, realizing the rotation \mbox{$R_Y \otimes R_Y$}. Plots of the
orbits and concurrence are shown in Fig.
\ref{fig:BELL_continuous_PRODUCT_momenta_R_i_R_i_sigma_1_2_4}.

Let us first consider $\sigma / m = 1$. At first, the effect of boosts is quite
similar to the previous case. When rapidity is smaller than $2.6$, the state is
mapped into a mixture of itself and the projector onto $\ket{\Psi_-}$, moving
along an orbit that connects the two states. At about $\xi = 2.6$, the boosted
observer sees a separable state. However, for larger boosts the orbit differs
from the previous case as the state moves back along the same path towards the
rest frame state. The concurrence mimics this pattern by first decreasing
monotonically until $\xi = 2.6$, and then increasing to almost maximal
entanglement for large boosts $\xi > 6$.

The orbits for $\sigma / m = 2$ and $\sigma / m = 4$ diverge from this
behavior, with the disagreement growing larger as the width increases. This is
to be expected since larger Gaussians contain spins some of which undergo less
and others more rotation than spins at the centre of the wave packet, thereby
causing the spin state to be a mixed state. Larger values of $\sigma / m$ lead
in general to a higher degree of mixedness of the boosted state, and the effect
becomes more pronounced at extremely large boosts: at $\xi = 6.5$, the boosted
state with $\sigma / m = 4$ is closer to the centre of the octahedron than the
states with lower $\sigma / m$.

\subsection{Case $R_i \otimes R_j$}\label{sec:bellContinuousMomentaR_i_R_j}

\begin{figure*}
	\centering
	\begin{subfigure}[t]{0.7\textwidth}
		\centering

\includegraphics[width=\textwidth]{%
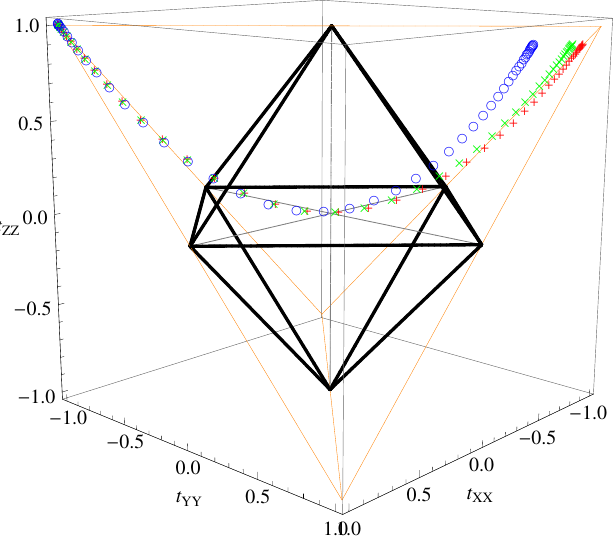}
		\caption{}
\label{fig:orbit_BELL_continuous_PRODUCT_momenta_R_i_R_k_sigma_1_2_4}
	\end{subfigure}
	\hspace{1em}
	\begin{subfigure}[t]{0.7\textwidth}
		\centering

\includegraphics[width=\textwidth]{%
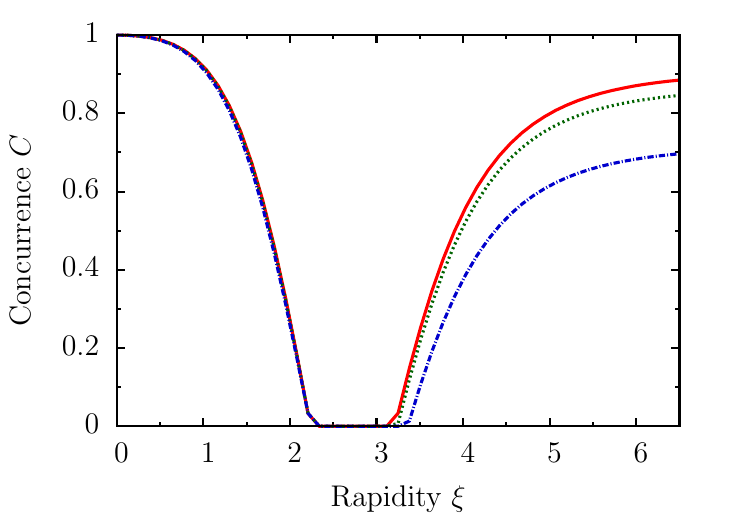}
		\caption{}
\label{fig:concurrence_BELL_continuous_PRODUCT_momenta_R_i_R_k_sigma_1_2_4}
	\end{subfigure}
	\caption%
[Spin (a) orbit and (b) concurrence under $R_i \otimes R_j$, $i \ne j$ for
Gaussian momenta $f_{\Sigma}$ with \mbox{$\sigma / m = 1, 2, 4$}.]%
{(Color online.) Spin (a) orbit and (b) concurrence under $R_i
\otimes R_j$, $i \ne j$ for Gaussian momenta with \mbox{$\sigma / m =
1, 2, 4$}. Product momenta are given by $f_{\Sigma}$. Data for
$\sigma / m = 1$ is shown with (a) the red ``+'' and (b) the red
solid line, $\sigma / m = 2$ with (a) the green ``$\times$'' and (b)
the green dotted line, $\sigma / m = 4$ with (a) the blue
``$\bigcirc$'' and (b) the blue dot-dashed line. (a)~Initial state
$\ket{\Phi_+}$ corresponds to vertex $(1, -1, 1)$.}
\label{fig:BELL_continuous_PRODUCT_momenta_R_i_R_k_sigma_1_2_4}
\end{figure*}

In order to realize scenarios where particles undergo rotations around different
axis,
the centers of Gaussians need to lie in different boost planes. With the boost
in the $z$-direction, we will choose \mbox{$\pm\ppp_0 = \pm\ppp_{Y0}$} and
\mbox{$\pm\qqq_0 = \pm\ppp_{X0}$}, which means that the spin state is rotated by
$R_X \otimes R_Y$. The orbits and concurrence are shown in
Fig.~\ref{fig:BELL_continuous_PRODUCT_momenta_R_i_R_k_sigma_1_2_4}.

Spin behavior under mixed rotations is quite different from the two previous
ones. Let us begin by considering $\sigma / m = 1$. The state follows an orbit
that has the shape of a curve starting at vertex $(1, -1, 1)$ and evolving
towards the origin, reaching it at about $\xi = 2.7$. The second half of the
orbit displays a symmetric shape. The state moves along a curve towards the
vertex $(-1, 1, 1)$ which represents the Bell state $\ket{\Phi_-}$, almost
reaching it when $\xi = 6.5$.

It is interesting that the spins become briefly separable in the interval $\xi
\in [2.2, 3.2]$. While this might look puzzling if one only had access to the
behavior of concurrence, the plot of orbits gives us deeper insight into what
is happening. The spin state evolves in the plane that intersects the
octahedron of separable states, entering the octahedron when $\xi = 2.2$ and
moving along a path towards the maximally mixed state $\frac{1}{4} \idd_4$
represented by $(0, 0, 0)$. At $\xi = 2.73$ the moving observer sees a
maximally mixed state. When the boosts become even larger, the entanglement
revives again, becoming non-zero for rapidities greater than $\xi = 3.2$, which
corresponds to the point where the state leaves the octahedron.

As above, we observe the generic feature that states with larger widths deviate
from this behavior at higher values of rapidity and the difference grows with
$\sigma / m$. While the orbits are fairly similar up to the maximally mixed
state, they start to diverge soon thereafter, with the momenta $\sigma / m = 4$
showing least gain in concurrence. Correspondingly, the latter state follows an
orbit in the set of states with lower degree of entanglement.

\section{Axis centered
Gaussians}\label{sec:twoParticlesContinuousOriginCentredGaussians}

\begin{figure}[htb]
\centering
\includegraphics[width=0.35\textwidth]
{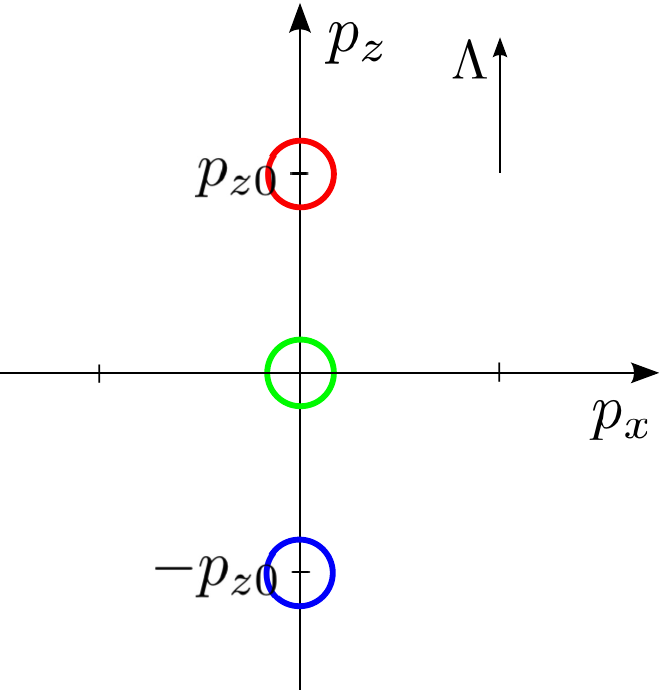}
\caption{(Color online.) Schematic illustration of axis centered
Gaussians shifted in the positive (top red circle) and negative
(bottom blue circle) direction, and centered at the origin (middle
green circle). Boost $\Lambda$ is in the positive $z$-direction.}
\label{fig:X6}
\end{figure}

One of the first studies of two particle entanglement in relativity was carried
through in the seminal paper \cite{gingrich_quantum_2002}, which focussed on
systems whose momenta were given by Gaussians centered at the origin. In this
section, we will study scenarios which are more general, involving momenta that
are centered on the $z$-axis. In particular, in the first scenario the Gaussian
momenta are shifted in the positive direction, $\ppp_0 = (0, 0, 4)$, in the
second in the negative direction, $\ppp_0 = (0, 0, -4)$, and in the third we
reproduce the origin centered momenta of \cite{gingrich_quantum_2002}, see
Fig.~\ref{fig:X6}. Fourthly, we will consider Gaussians that are far away from
the origin, $\ppp_0 = (0, 0, -98.5)$, and thus likely to induce large rotations
on the spins.

\begin{figure*}
	\centering
	\begin{subfigure}[t]{0.7\textwidth}
		\centering

\includegraphics[width=\textwidth]{%
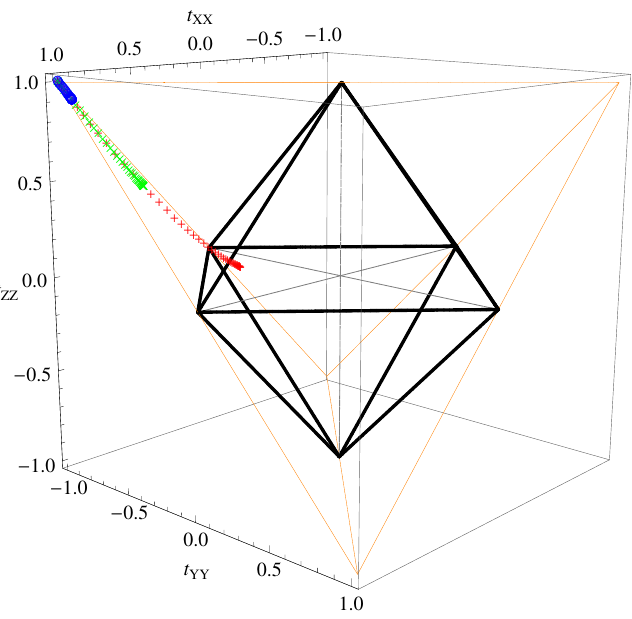}
		\caption{}

\label{fig:orbit_BELL_continuous_PRODUCT_momenta_R_i_R_k_sigma_1_2_4}
	\end{subfigure}
	\hspace{1em}
	\begin{subfigure}[t]{0.7\textwidth}
		\centering

\includegraphics[width=\textwidth]{%
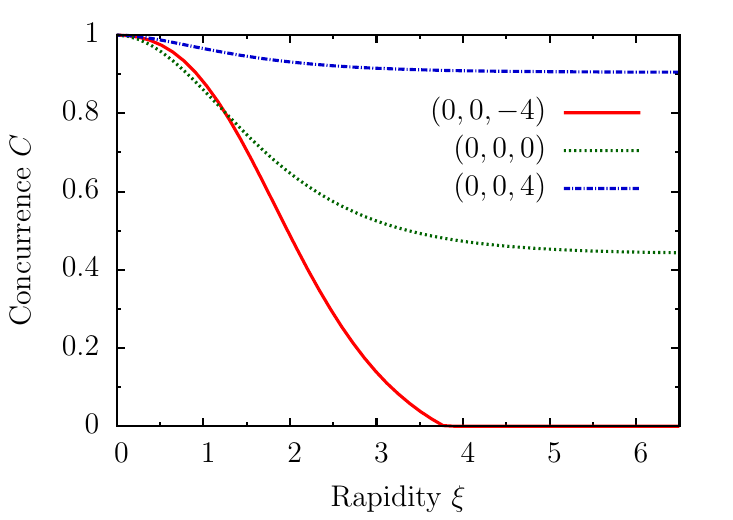}
		\caption{}
	\label{fig:concurrence_BELL_continuous_PRODUCT_momenta_GA_sigma1_-4-0-4}
	\end{subfigure}
	\caption%
[Spin (a) orbit and (b) concurrence for origin centered Gaussian momenta with
$\sigma / m = 1$.]%
{(Color online.)
Spin (a) orbit and (b) concurrence for origin centered Gaussian
momenta with $\sigma / m = 1$. Data for $(0, 0, -4)$ is shown with
(a) the red ``+'' and (b) the red solid line, $(0, 0, 0)$ with (a)
the green ``$\times$'' and (b) the green dotted line, $(0, 0, 4)$
with (a) the blue ``$\bigcirc$'' and (b) the blue dot-dashed line.
(a)~Initial state $\ket{\Phi_+}$ corresponds to vertex $(1, -1, 1)$.}
	\label{fig:BELL_continuous_PRODUCT_momenta_GA_sigma1_-4-0-4}
\end{figure*}

\begin{figure*}
	\centering
	\begin{subfigure}[t]{0.7\textwidth}
		\centering

\includegraphics[width=\textwidth]{%
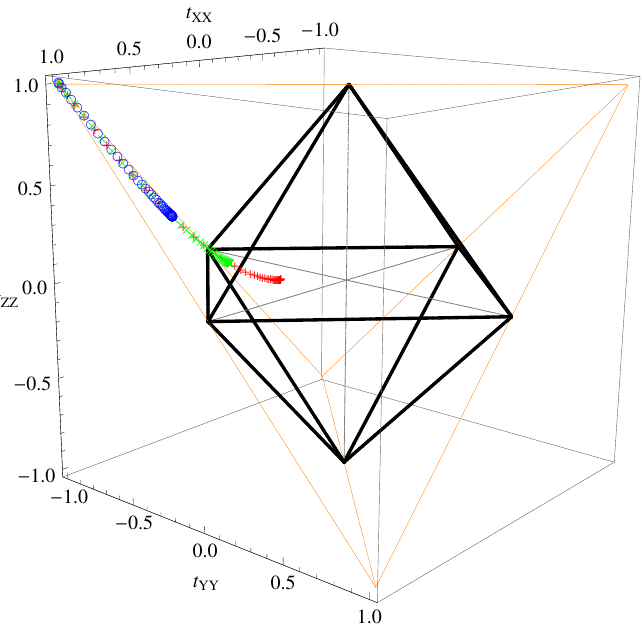}
		\caption{}
\label{
fig:orbit_BELL_continuous_PRODUCT_momenta_GA_state_sigma4_mathematica_-4-0-4}
	\end{subfigure}
	\hspace{1em}
	\begin{subfigure}[t]{0.7\textwidth}
		\centering

\includegraphics[width=\textwidth]{%
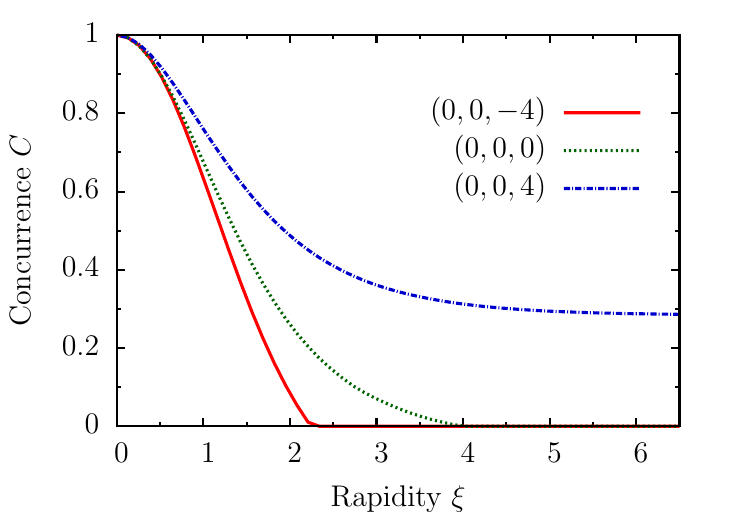}
		\caption{}
\label{fig:concurrence_BELL_continuous_PRODUCT_momenta_GA_state_sigma4_-4-0-4}
	\end{subfigure}
	\caption%
[Spin (a) orbit and (b) concurrence for origin centered Gaussian momenta with
$\sigma / m = 4$.]%
{(Color online.)
Spin (a) orbit and (b) concurrence for origin centered Gaussian
momenta with $\sigma / m = 4$. Data for $(0, 0, -4)$ is shown with
(a) the red ``+'' and (b) the red solid line, $(0, 0, 0)$ with (a)
the green ``$\times$'' and (b) the green dotted line, $(0, 0, 4)$
with (a) the blue ``$\bigcirc$'' and (b) the blue dot-dashed line.
(a)~Initial state $\ket{\Phi_+}$ corresponds to vertex $(1, -1, 1)$.}
	\label{fig:BELL_continuous_PRODUCT_momenta_GA_sigma4_-4-0-4}
\end{figure*}

Plots for \mbox{$\sigma / m = 1$} and \mbox{$\sigma / m = 4$} with the first
three momenta are shown in
Fig.~\ref{fig:BELL_continuous_PRODUCT_momenta_GA_sigma1_-4-0-4} and
\ref{fig:BELL_continuous_PRODUCT_momenta_GA_sigma4_-4-0-4}. The results of
\cite{gingrich_quantum_2002} correspond to the Gaussian momenta which have
$\sigma / m = 1$ and $\sigma / m = 4$ and where the momenta are centered at
$(0, 0, 0)$.

Let us consider first $\sigma / m = 1$. Note that we have changed tack a
little. Whereas in the previous sections we kept the center of the Gaussian
momentum fixed, here we keep its width fixed and change the coordinate of the
center. The differences between the three scenarios in
Fig.~\ref{fig:BELL_continuous_PRODUCT_momenta_GA_sigma1_-4-0-4} are quite
dramatic. While $\ppp_0 = (0, 0, 4)$ shows relatively little decrease of
entanglement with the concurrence saturating at $0.9$ for large boosts $\xi =
6.5$, the system with $\ppp_0 = (0, 0, 0)$ loses more than half of the
entanglement and saturates at $0.45$. The third one with momentum at $\ppp_0 =
(0, 0, -4)$ displays a steep decrease of concurrence, with the entanglement
vanishing altogether for rapidities $\xi > 3.75$. Momenta with $\sigma / m = 4$
exhibit similar features, albeit with much steeper decreases of concurrence.
Even for $\ppp_0 = (0, 0, 4)$, the boosted state has only about $0.3$ of the
original degree of entanglement at large boosts, and $\ppp_0 = (0, 0, -4)$
vanishes already at $\xi = 2.2$. The corresponding orbits follow a trajectory
which evolve towards the base of the upper pyramid of separable states, see
Figs.~\ref{fig:orbit_BELL_continuous_PRODUCT_momenta_R_i_R_k_sigma_1_2_4} and
\ref{
fig:orbit_BELL_continuous_PRODUCT_momenta_GA_state_sigma4_mathematica_-4-0-4}.
All the orbits follow the same path, the only difference lying in that some
stop sooner than others. The latter is determined by the location and width of
the Gaussian. Momenta whose centers are shifted farther in the negative
direction and have larger widths correspond to the final states closer to the
center of the octahedron and exhibit, consequently, a quicker and steeper
decline of the concurrence.

\subsection{Comparison with single
particle}\label{sec:ComparisonWithSingleParticle}

It is instructive to discuss how these results relate to a single particle
system with axis centered momenta \cite{palge_generation_2012}. At first sight
it might seem that the single and two particle systems are not directly
comparable because the entanglements in question are between different degrees
of freedom: spin--momentum entanglement in case of the single particle versus
spin--spin in case of two particles. Correspondingly, the structure of maps that
change entanglement in each case as well as the initial states of systems are
different too. However, despite this a number of analogies are manifest and we
will argue that this is no coincidence. Both systems show features which can be
explained using the properties of TWR.

Firstly, the two particle scenario with $\ppp_0 = (0, 0, 4)$, which involves
momenta in the direction of boost, displays less pronounced changes of
entanglement than the one with a Gaussian centered at $\ppp_0 = (0, 0, -4)$,
which has momenta opposite to the direction of boost. As discussed in
\cite{palge_generation_2012}, this originates in the sensitivity of TWR to the
angle between boosts. Smaller boost angles lead to smaller TWR, which in turn
result in smaller changes of entanglement. Secondly, in analogy with the single
particle, Gaussians with larger widths show in general more rapid changes of
concurrence. This can be traced back to the dependence of TWR on the magnitude
of the boost. A Gaussian with a larger width is equivalent to a system
undergoing a larger boost, which in turn causes a larger TWR angle. Thirdly,
both single and two particle systems exhibit saturation, which comes from the
fact that the TWR achieves a maximum value, which for a given boost angle is
determined by the smaller boost.

\subsection{Large momenta}

\begin{figure*}
	\centering
	\begin{subfigure}[t]{0.7\textwidth}
		\centering

\includegraphics[width=\textwidth]{%
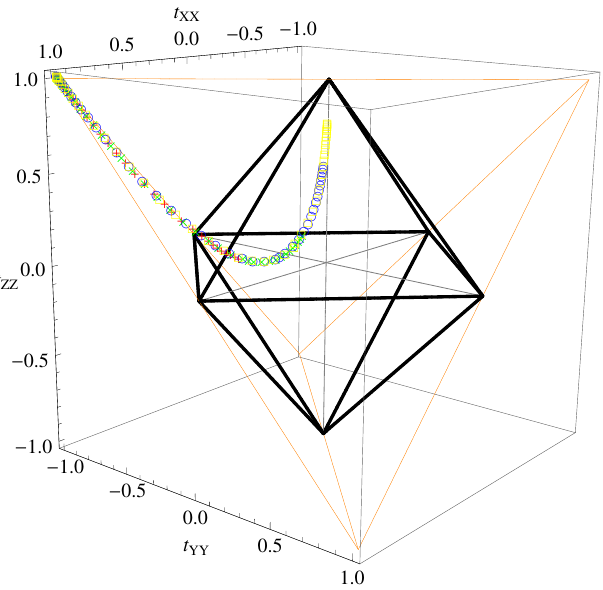}
		\caption{}
\label{
orbit_BELL_continuous_PRODUCT_momenta_GA_state_EXTREME_rotation_mathematica_sigm
a_1-2-4_0-0--98}
	\end{subfigure}
	\hspace{1em}
	\begin{subfigure}[t]{0.7\textwidth}
		\centering

\includegraphics[width=\textwidth]{%
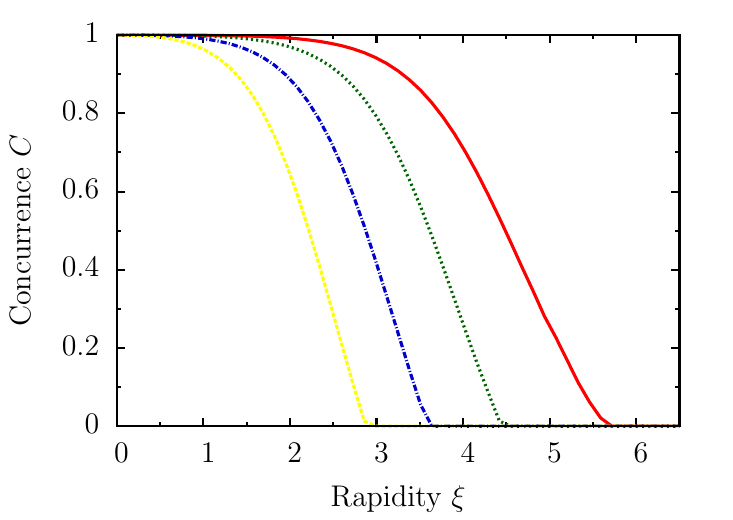}
		\caption{}
\label{
fig:concurrence_BELL_continuous_PRODUCT_momenta_GA_state_EXTREME_rotation_sigma_
1_2_4_0-0--98}
	\end{subfigure}
	\caption%
[Spin (a) orbit and (b) concurrence for Gaussian momenta $f_C$ with
\mbox{$\sigma / m = 1, 2, 4, 8$} and $\ppp_0 = (0, 0, -98.5)$.]%
{(Color online.) Spin (a) orbit and (b) concurrence for axis centered
Gaussian momenta $f_C$ with \mbox{$\sigma / m = 1, 2, 4, 8$} and
$\ppp_0 = (0, 0, -98.5)$. Data for $\sigma / m = 1$ is shown with (a)
the red ``+'' and (b) the red solid line, $\sigma / m = 2$ with (a)
the green ``$\times$'' and (b) the green dotted line, $\sigma / m =
4$ with (a) the blue ``$\bigcirc$'' and (b) the blue dot-dashed line,
$\sigma / m = 8$ with (a) the yellow ``$\Box$'' and (b) the yellow
dotted leftmost line.
(a)~Initial state $\ket{\Phi_+}$ corresponds to vertex $(1, -1, 1)$.}
\label{fig:BELL_continuous_PRODUCT_momenta_GA_state_EXTREME_rotation_0-0--98}
\end{figure*}

\begin{figure}[htb]
\centering
\includegraphics[width=0.35\textwidth]{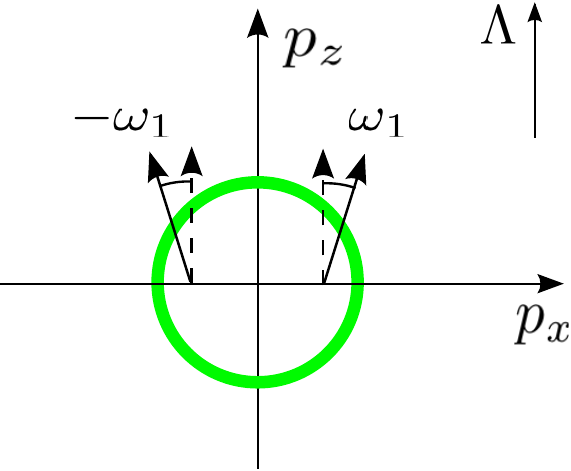}
\caption%
[Schematic representation of an origin centered Gaussian spin field.]
{Schematic representation of an origin centered Gaussian spin field.
}%
\label{fig:SchematicRepresentationOriginCentredGaussianSpinField}
\end{figure}

Let us next turn to the case of large momenta \mbox{$\ppp_0 = (0, 0, -98.5)$}.
Plots for \mbox{$\sigma / m = 1, 2, 4, 8$} are shown in
Fig.~\ref{fig:BELL_continuous_PRODUCT_momenta_GA_state_EXTREME_rotation_0-0--98}.
Interestingly, and contrary to what one might expect based on the findings so
far, entanglement declines more slowly than in the previous scenarios. For
instance, states with $\sigma / m = 1$ remain nearly maximally entangled for
rapidities up to about $2$ and decohere thereafter, but this occurs later than
with the momenta $\ppp_0 = (0, 0, -4)$, which on the face of it generate smaller
rotation angles than the extreme momenta $\ppp_0 = (0, 0, -98.5)$. However, on
closer examination such puzzling behavior can be again explained using the
properties of TWR. Instead of a Gaussian, let us think of a rough, simple model
consisting of discrete momenta in the \plane{x}{z} as depicted in
Fig.~\ref{fig:SchematicRepresentationOriginCentredGaussianSpinField}. We know
that larger momenta generate larger rotation angles, but their amplitude is
smaller, so for the sake of argument, let us assume that the Gaussian is
represented by two momenta at the distance of $0.75$ its width.
We will next argue that concurrence changes more rapidly for the Gaussian
centered at or close to the origin than for the one centered at the very large
momentum $(0, 0, -98.5)$. The key is to realize that the boost angle $\theta$
is $\pi / 2$ for the origin centered Gaussian, while it is larger, about
$170\degree$ or $2.97$ rad, for the Gaussian at $(0, 0, -98.5)$. In
Fig.~\ref{fig:figureTWRFunctionOfRapidityBoostAngle}, which describes the
dependence of TWR angle on boost angle and rapidity, these states lie,
respectively, in the middle and almost at the right end of the horizontal axis.
Boosting the system means we keep $\theta$ fixed and move towards the back of
the surface representing the TWR angle for the given $\theta$ and $\xi$. Now
for $\theta = \pi / 2$, the rotation grows initially faster than for $\theta =
2.85$, meaning that the concurrence of the origin centered Gaussian changes
sooner than the one at the extremely large momentum. However, as rapidity grows
even larger, the rotation increases rapidly for $\theta = 2.85$, leading to the
decrease of concurrence as seen in
Fig.~\ref{fig:BELL_continuous_PRODUCT_momenta_GA_state_EXTREME_rotation_0-0--98}.
The decrease becomes steeper as width increases, as is to be expected
since larger width means we move towards slightly lower values of
$\theta$ in Fig.~\ref{fig:figureTWRFunctionOfRapidityBoostAngle}
which cause faster rotations and hence quicker drop of concurrence.
Along the same lines, for Gaussians at $(0, 0, -4)$ which are
relatively close to the origin in comparison to $(0, 0, -98.5)$,
$\theta$ is slightly but not significantly larger than $\pi / 2$,
still leading to faster initial increase than for the extremely large
momenta.

\begin{figure}[htb]
\centering
\includegraphics[width=0.5\textwidth]{%
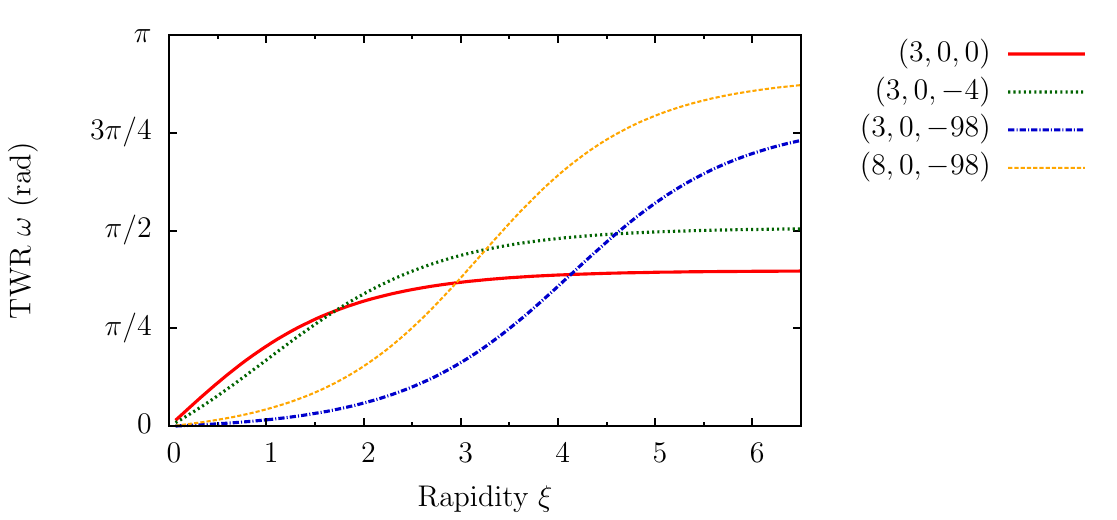}
\caption%
[TWR rotation for axis centered Gaussians in different geometries.]
{(Color online.) TWR for axis centered Gaussians in different geometries.
}%
\label{fig:twoParticlesExplainingAwayPuzzlingOriginCentredBehaviour}
\end{figure}

To substantiate these qualitative considerations with a rough numerical model,
we plot the dependence of TWR on rapidity for four delta momenta in
Fig.~\ref{fig:twoParticlesExplainingAwayPuzzlingOriginCentredBehaviour}.
The first one at $(3, 0, 0)$ corresponds to the origin centered Gaussian and the
second $(3, 0, -4)$ to the one close to the origin. The third $(3, 0, -98)$
represents a distribution with the same width at the extreme momentum and the
fourth $(8, 0, -98)$ corresponds to a Gaussian with larger width at the extreme
momentum. The qualitative behavior of TWR and hence of concurrence follows the
pattern we have just outlined. Quantitatively, however, our discrete
considerations in the 2D setting cannot accurately represent the more complex
workings of realistic 3D Gaussian wave packets. The model in
Fig.~\ref{fig:twoParticlesExplainingAwayPuzzlingOriginCentredBehaviour} does not
reproduce the precise numerical values for concurrence in
Figs.~\ref{fig:BELL_continuous_PRODUCT_momenta_GA_sigma1_-4-0-4},
\ref{fig:BELL_continuous_PRODUCT_momenta_GA_sigma4_-4-0-4}
and~\ref{fig:BELL_continuous_PRODUCT_momenta_GA_state_EXTREME_rotation_0-0--98}.

To summarize, the claim we make is that the behavior of a Gaussian system can be
understood qualitatively, and to some extent even quantitatively, using a rather
simple model involving a small sample of discrete (or very narrow Gaussian)
momenta.

\section{Product momenta $f_{\times}$}\label{sec:productMomentaFTimesGA}

We will next study product momenta of the form $f_{\times}$. Above we introduced
them as a generalization of $f_{\Sigma}$. In this section, however, we will show
that they serve another purpose as well: in many cases, they can be used to
model the axis centered Gaussians of the previous section. This has mainly the
conceptual importance of providing a rough and ready explanation of how the axis
centered systems behave. The practical use of this exercise is somewhat limited
since we will not provide systematic methods for finding the exact parameters
that characterize such models.

We start by noting that the state $f_{\times}$ admits only two types of
rotations: $R_i \otimes \idd$ and a mixture of $R_i \otimes R_i$ and $R_i
\otimes R_j$. We will forgo the former type since it is not interesting from the
point of view of comparison with the axis centered systems. To discuss the
latter type, we will first consider the case where the geometric vectors of
$f_{\times}$ are described by the large momenta
\begin{align}
\pm\ppp_0 = \pm\qqq_0 = \pm\ppp_{Y0}, \quad \pm\ppp^{\perp}_0 =
\pm\qqq^{\perp}_0 = \pm\ppp_{X0}, \nonumber
\end{align}
which guarantee that spins undergo almost maximum TWRs. Plots of the spin orbits
and concurrence for $\sigma / m = 1, 2, 4$ are shown in
Fig.~\ref{
fig:BELL_continuous_PRODUCT_momenta_FSIGMA_mathematica_sigma_1_2_4_17-17--98}.
\begin{figure*}
	\centering
	\begin{subfigure}[t]{0.7\textwidth}
		\centering

\includegraphics[width=\textwidth]{%
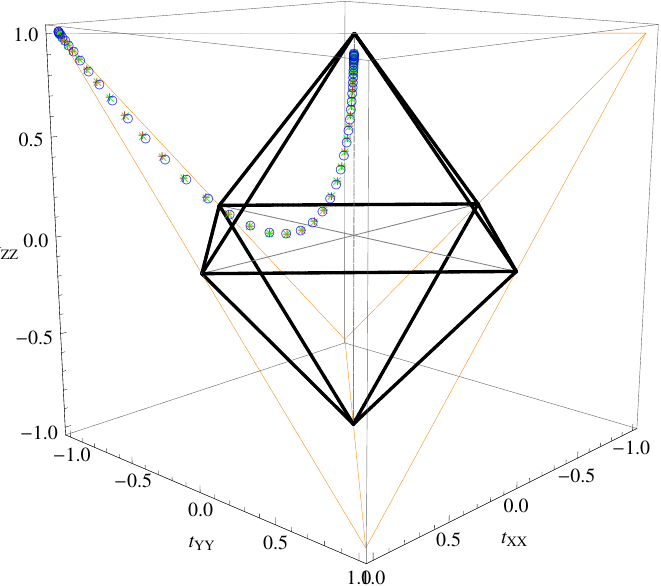}
		\caption{}
\label{
fig:orbit_BELL_continuous_PRODUCT_momenta_FSIGMA_mathematica_sigma_1_2_4_17-17--98}
	\end{subfigure}
	\hspace{1em}
	\begin{subfigure}[t]{0.7\textwidth}
		\centering

\includegraphics[width=\textwidth]{%
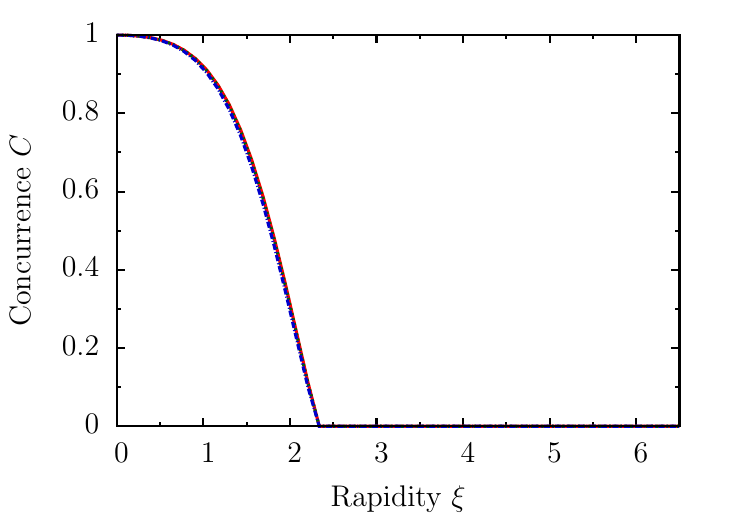}
		\caption{}
\label{fig:concurrence_BELL_continuous_PRODUCT_momenta_R_i_R_k_ga_sigma_1_2_4}
	\end{subfigure}
	\caption%
[Spin (a) orbit and (b) concurrence for momenta $f_{\times}$ with \mbox{$\sigma
/ m = 1, 2, 4$}.]%
{(Color online.) Spin (a) orbit and (b) concurrence for Gaussian
momenta with \mbox{$\sigma / m = 1, 2, 4$}. Product momenta are given
by $f_{\times}$ with $\pm\ppp_0 = \pm\ppp_{X0}$ and
$\pm\ppp^{\perp}_0 = \pm\ppp_{Y0}$. Data for $\sigma / m = 1$ is
shown with (a) the red ``+'' and (b) the red solid line, $\sigma / m
= 2$ with (a) the green ``$\times$'' and (b) the green dotted line,
$\sigma / m = 4$ with (a) the blue ``$\bigcirc$'' and (b) the blue
dot-dashed line. (a)~Initial state $\ket{\Phi_+}$ corresponds to
vertex $(1, -1, 1)$.}
\label{
fig:BELL_continuous_PRODUCT_momenta_FSIGMA_mathematica_sigma_1_2_4_17-17--98}
\end{figure*}

The orbits exhibit interesting behavior, initially showing a pattern that is
analogous to the state $f_{\Sigma}$ for the case $R_i \otimes R_j$, see
Fig.~\ref{fig:BELL_continuous_PRODUCT_momenta_R_i_R_k_sigma_1_2_4}. However,
after arriving the octahedron, we see different behavior: the orbit changes
course and evolves towards the top of the upper pyramid. When the spins reach
maximal rotation, the state becomes close to an equal mixture of projectors onto
$\ket{\Phi_+}$ and $\ket{\Phi_-}$, never leaving the octahedron of separable
states. This explains why concurrence vanishes for all $\xi > 2.3$.

Let us next consider the correspondence between the $z$-axis centered momenta
and the $f_{\times}$ model. When analyzing the curious behavior of the $z$-axis
Gaussians in the previous section, we resorted to a naive 2D model in the
\plane{x}{z}. Realistic Gaussians however involve a third dimension as well, and
generalizing the 2D model to three dimensions naturally leads to the state which
is given by $f_{\times}$. This explains why there is a close match between the
orbits of the $z$-axis centered states with the large momenta $(0, 0, -98.5)$
and those of $f_{\times}$ above. This raises the question of whether the
$z$-axis centered states shown in
Fig.~\ref{fig:BELL_continuous_PRODUCT_momenta_GA_sigma4_-4-0-4} can be modeled
using the $f_{\times}$ states with suitably chosen momenta. Proceeding in the
same naive way as for the 2D model, let us approximate the states in
Fig.~\ref{fig:BELL_continuous_PRODUCT_momenta_GA_sigma4_-4-0-4} using
$f_{\times}$ and assuming that the momenta are described by
\begin{align}
\pm\ppp_0 = \pm\qqq_0 = (0, \pm 3, p_z), \quad \pm\ppp^{\perp}_0 =
\pm\qqq^{\perp}_0 = (\pm 3, 0, p_z), \nonumber
\end{align}
where $p_z$ takes the values $-4$, $0$ and $4$. We plot the orbits and
concurrence for $\sigma / m = 0.25$ in
Fig.~\ref{
fig:BELL_continuous_PRODUCT_momenta_FSIGMA_mathematica_sigma_0-25_3-3--4_0_4},
where we have chosen $\sigma$ to be smaller than above in order to minimize
width related effects.
\begin{figure*}
	\centering
	\begin{subfigure}[t]{0.7\textwidth}
	\centering

\includegraphics[width=\textwidth]{%
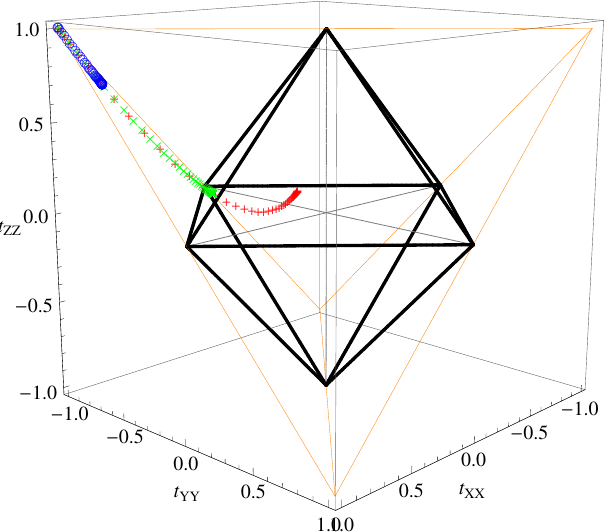}
		\caption{}
\label{
fig:orbit_BELL_continuous_PRODUCT_momenta_FSIGMA_mathematica_sigma_0-25_3-3--4_0_4}
	\end{subfigure}
	\hspace{1em}
	\begin{subfigure}[t]{0.7\textwidth}
		\centering

\includegraphics[width=\textwidth]{%
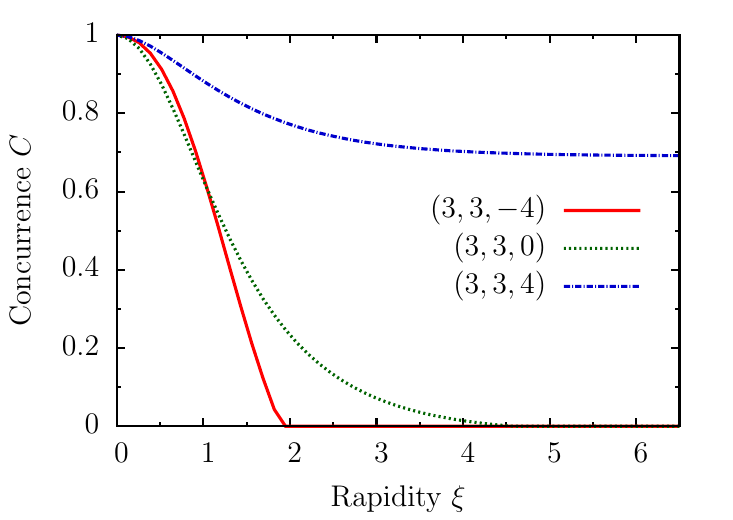}
		\caption{}
\label{
fig:concurrence_BELL_continuous_PRODUCT_momenta_FTIMES_sigma_0-25_3-3--4_0_4}
	\end{subfigure}
	\caption%
[Spin (a) orbit and (b) concurrence for momenta $f_{\times}$ with \mbox{$\sigma
/ m = 0.25$}.]%
{(Color online.) Spin (a) orbit and (b) concurrence for Gaussian
momenta with \mbox{$\sigma / m = 0.25$}. Product momenta are given by
$f_{\times}$. Data for $(3, 3, -4)$ is shown with (a) the red ``+''
and (b) the red solid line, $(3, 3, 0)$ with (a) the green
``$\times$'' and (b) the green dotted line, $(3, 3, 4)$ with (a) the
blue ``$\bigcirc$'' and (b) the blue dot-dashed line. (a)~Initial
state $\ket{\Phi_+}$ corresponds to vertex $(1, -1, 1)$.}
\label{
fig:BELL_continuous_PRODUCT_momenta_FSIGMA_mathematica_sigma_0-25_3-3--4_0_4}
\end{figure*}

While the agreement with
Fig.~\ref{fig:BELL_continuous_PRODUCT_momenta_GA_sigma4_-4-0-4} is not perfect,
one can easily recognize the features present in the original $z$-axis case. The
concurrence of the $f_{\times}$ model exhibits roughly the same kind of
dependence on the boost angle as the $z$-axis centered states. Although the
momenta with $p_z = 4$ diverge considerably from those with $(0, 0, 4)$ in
Fig.~\ref{fig:BELL_continuous_PRODUCT_momenta_GA_sigma4_-4-0-4}, the fit is
relatively good for $p_z = -4$ and $p_z = 0$ considering this is a simple model.
The orbits follow the same pattern, with the one for $p_z = 4$ deviating more,
and those for $p_z = 0$ and $p_z = -4$ relatively little from the $z$-axis
centered states.

To summarize, all along we have been using the notion that systems involving
continuous momenta, and specifically those of Gaussian form, can be understood
in terms of discrete models, possibly containing many momentum eigenstates. The
foregoing discussion bolsters this claim by showing that in some cases Gaussian
momenta admit very simple models. In particular the momenta centered at the
axis parallel to the direction of boost can be modeled by sampling four narrow
Gaussians.

\section{Correspondence to discrete systems}

We would like to comment on the relation between continuous and discrete systems
which is implicit in all the cases discussed above: when the width of the
Gaussian becomes small enough, we observe a good match with discrete systems. In
many cases, the behavior of the latter can be calculated analytically
\cite{palge_behavior_2015}.

By way of example, consider rotations of type $R_i \otimes R_i$ generated by
product momenta $f_{\Sigma}$. Comparison with the plots of the discrete model,
see Fig.~5 in \cite{palge_behavior_2015}, shows that for $\sigma / m = 1$ the
behavior of the continuous and the discrete model coincide to quite a high
degree of accuracy. The orbit of the continuous system follows the same path as
the discrete one, almost reaching the rest frame state $\ket{\Phi_+}$. The
reason it stops short of $\ket{\Phi_+}$ is that while in the discrete model we
assume that the system reaches the maximum TWR of $180\degree$, the maximum
rotation implemented by the continuous model at $\xi = 6.5$ is $\omega_m \approx
163\degree$ or $2.81\;\text{rad}$. Substituting $\omega_m$ into the expression
that describes the discrete orbit, Eq.~(63) in \cite{palge_behavior_2015},
yields \mbox{$t_{Y \otimes Y}(\omega_m) = (0.9, -1, 0.9)$}, which is in good
agreement with the numerically calculated value $(0.89, -0.99, 0.90)$
representing the final state for \mbox{$\sigma / m = 1$} in
Fig.~\ref{fig:orbit_BELL_continuous_PRODUCT_momenta_R_i_R_i_sigma_1_2_4}.
Likewise, the concurrence of the discrete model, Eq.~(62) with $\lambda = 1$ in
\cite{palge_behavior_2015}, evaluates to \mbox{$C(\omega_m) = 0.89$}, showing
again good fit with the continuous model.

This pattern is generic in that a similar analysis can be run for each type of
rotation. Although it might seem that the case $R_i \otimes \idd$ in section
\ref{sec:bellContinuousMomentaR_i_1} provides a counterexample, this is not
true. The reason it deviates from the discrete behavior is that the identity map
can not be implemented accurately enough. Realistic systems that are
characterized by wave packets of finite width always contain momenta which
induce some rotation on the spin field, thereby diverging from idealized
behavior.

\section{Discussion and summary}\label{sec:conclusion}

In this paper we have studied spin entanglement of two particles with
continuous momenta. We have surveyed a number of boost scenarios involving
momenta in product states. Attention was confined to pure spins, which were
assumed to be in the maximally entangled Bell state $\ket{\Phi_+}$.

Our results confirm the general conclusion that Lorentz boosts cause
non-trivial behavior of spin entanglement of a two particle system. The details
of the behavior, however, are strongly determined by the boost situation at
hand, that is, the momentum state and geometry involved. While there are states
and geometries that leave entanglement invariant, most scenarios we have
studied lead to significant changes of concurrence. An example of the former
was given by the product momenta $f_{\mathrm{EPRB}}$ with $\sigma / m = 1$ in
the EPRB situation which leaves the entanglement of the Bell state invariant.
The rest of the momenta causes changes of spin entanglement between the maximal
value and zero.

Although the analysis was numerical throughout, the lack of analytic models was
to some extent compensated by modeling continuous momenta in terms of discrete
ones. In this picture, systems involving continuous momenta can be thought of
as fields comprising spins at a large number of discrete momenta, where
boosting means that each spin undergoes a different, momentum dependent
rotation for a given value of rapidity. The difference between the behaviors of
the EPRB momenta and the rest of the systems can then be explained in terms of
the rotations that the discrete models generate on the spin degree of freedom.

It is also worthwhile highlighting the different roles that momentum states and
geometries play in a boost scenario. Fixing a momentum state is equivalent to
choosing a particular class of spin orbits from the set of all possible orbits.
The boost geometry, on the other hand, gives a handle that enables one to tune
the  magnitude of the rotation that the spins are subjected to. In other words,
specifying a geometry means picking a particular spin orbit from the class of
spin orbits associated with a certain momentum state. As an illustration,
consider
Fig.~\ref{
fig:BELL_continuous_PRODUCT_momenta_FSIGMA_mathematica_sigma_0-25_3-3--4_0_4}
which shows the same momentum state $f_{\times}$ with three different boost
angles. By specifying that the momenta are given by $f_{\times}$ we determine
that the spin orbit is the one associated with the momentum $f_{\times}$ as
opposed to, for instance, $f_{\Sigma}$. Further, by fixing the boost angle one
determines the upper bound of the TWR for spins, thereby choosing a particular
orbit from the class associated with $f_{\times}$. The reason for choosing
extremely large momenta and large boost angles was that we wanted to obtain the
longest orbit in the particular class. Scenarios with smaller boost angles are
subsumed in the sense that they are given by shorter orbits in the same class:
orbits whose endpoint corresponds to a smaller maximum TWR.

We would also like to comment on the role of the initial states. While we
assumed from the start that the focus is exclusively on systems whose spin and
momentum degrees of freedom factorize, the spin--momentum entangled states have
been, to some extent, implicit in the investigation too. This is because all
inertial frames are equivalent and Lorentz boosts are group elements, meaning
that we are guaranteed to have inverse elements and the scenarios can be read
in the reverse direction. One can regard the final state, which typically
contains spin--momentum entanglement, as the rest frame state, and take the
inverse boost to obtain the initial state. For instance, consider the boosted
state $\frac{1}{4} \idd_4$ at $\xi = 2.2$, which is represented by $(0, 0, 0)$
in Fig.~\ref{fig:BELL_continuous_PRODUCT_momenta_R_i_R_k_sigma_1_2_4}. Applying
the inverse boost gives back the original maximally entangled Bell state
$\ket{\Phi_+}$. All plots can be interpreted this way.

This points to an important asymmetry between spin--momentum product versus
entangled states. Whereas the latter can lead to an increase of spin--spin
entanglement, it has been shown that the former can never cause such behavior
\cite{gingrich_quantum_2002}.

Finally, we would like to emphasize the usefulness of visualization of spin
orbits, which provided further insight into the behavior of entanglement. We
gained a more detailed understanding of how varying the initial states, their
widths and momenta, changed the spin concurrence. The hope is that the results
obtained in this paper contribute to a better understanding of entanglement in
relativity and could lead to future applications which might be of interest in
relativistic quantum information.

\section{Acknowledgments}

Veiko Palge was supported by EU through the ERDF CoE program grant TK133 and by
the Estonian Research Council via IUT2-27. Stefan Groote and Hannes Liivat were
supported by the Estonian Research Council via IUT2-27. We would like to thank
Hardi Veerm{\"a}e for a number of helpful suggestions.

\appendix

\section{particles in the Wigner representation}\label{sec:AppendixA}

\subsection{Conventions}

We will use natural units where $\hbar = c = 1$. Spacetime metric is
$\mathrm{diag}(+ - - -)$. Latin indices $i, j, k$ etc.\ take values in three
tuples $(x, y, z)$ or $(1, 2, 3)$ while Greek indices $\mu, \nu$ etc.\ run over
$(t, x, y, z)$ or $(0, 1, 2, 3)$. Three vectors use boldface whereas four
vectors are given in ordinary type. For instance, the four momentum is $p^{\mu}
= (p^0, \mathbf{p})$ with the norm $p^{\mu} p_{\mu} = (p^0)^2 - \mathbf{p}^2 =
m^2$.

\subsection{Particles}

In this section, we summarize the background for the relativistic quantum
mechanical constructions used in the paper. Throughout we work in the Wigner
representation which can be found in references
\cite{bogolubov_introduction_1975, sexl_relativity_2001}. The single particle
states are given by a unitary irreducible representation of the Poincar\'{e}
group where a representation is labelled by mass $m > 0$ and the intrinsic spin
$s$ which takes integral or half-integral values. The representation can be
realized in the space $\bigoplus^{2s + 1} L^2(\Gamma^+_m)$ of square integrable
functions on the forward mass hyperboloid $\Gamma^+_m = \{ \pppp \in \mink:
\pppp^2 = m^2, \pppp^0 > 0 \}$ where the scalar product is defined as
\begin{align}
\braket{\phi}{\psi} = \sum_\sigma \int \dddd\mu(p)\, \phi^*_{\sigma}(p)
\psi_{\sigma}(p),
\end{align}
with $\dddd\mu(p) = [2E(\ppp)]^{-1}\dddd^3\ppp$ being the Lorentz invariant
integration measure. In this paper we specialize on spin-$1/2$ systems, then
the state space is given by
\begin{align}
\hilb = L^2(\RRR) \oplus L^2(\RRR) = L^2(\RRR, \CC) = L^2(\RRR) \otimes \CC.
\end{align}
In order to define basis vectors, we start by specifying
the rest frame states in terms of four momentum $P^{\mu}$, square of total
angular momentum $\mathbf{J}^2$ and the $z$-component of angular momentum
$J_z$,
\begin{align}
P^{\mu} \ket{\mathbf{0}, \lambda} &= p^\mu_0 \ket{\mathbf{0}, \lambda}, \nonumber\\
\mathbf{J}^2 \ket{\mathbf{0}, \lambda} &= s (s + 1) \ket{\mathbf{0}, \lambda}, \\
J_z \ket{\mathbf{0}, \lambda} &= \lambda \ket{\mathbf{0}, \lambda},\nonumber
\end{align}
where $\mbf{0}$ denotes $\ppp = 0$ with $p^\mu_0 = (m, \mbf{0})$, and we have
abbreviated $\ket{\ppp, \lambda} = \ket{\ppp} \otimes \ket{\lambda}$. Because
the particle is at rest, $s$ and $\lambda$ refer to the spin and the
$z$-component of the particle. We next generate a complete basis, which
consists of the general eigenvectors of $P^{\mu}$, by acting on the rest frame
state with a pure, rotation free Lorentz boost,
\begin{align}
\ket{\ppp, \lambda} = U[L(\ppp)] \ket{\mbf{0}, \lambda},
\end{align}
where $U[L(\ppp)]$ is a unitary representation of boost $L(\ppp)$ that takes
the rest momentum $(m, \mbf{0}) = p_0$ to an arbitrary momentum,
\begin{align}
L(\ppp)\, (m, \mbf{0}) = (E(\ppp), \ppp),
\end{align}
with $E(\ppp) = \sqrt{\ppp^2 + m^2}$. The basis vectors $\ket{\ppp, \lambda}$
span the single particle state space $\mathcal{H}$ and we can write a generic
state as
\begin{align}
\ket{\Psi} = \sum_{\sigma} \int \dddd\mu(p)\, \psi_\sigma(\ppp) \ket{\ppp, \sigma},
\end{align}
The basis states are normalized as follows,
\begin{align}
\langle \ppp', \sigma' | \ppp, \sigma \rangle = 2 E(\ppp) \delta^3(\ppp -
\ppp') \delta_{\sigma \sigma'}.
\end{align}
The action of a generic Lorentz transformation $\Lambda$ on an element of basis
is given by
\begin{align}
U(\Lambda) \ket{\ppp, \sigma} = \sum_{\lambda} \ket{\Lambda\ppp, \lambda}
D_{\lambda\sigma}[W(\Lambda, \ppp)],
\end{align}
where $W(\Lambda, \ppp)$ is the Wigner rotation
\begin{align}
W(\Lambda, \ppp) \equiv L^{-1}(\Lambda\ppp) \Lambda L(\ppp)
\end{align}
that leaves $p_0$ invariant, $p_0 = W p_0$. For massive particles, $W \in
\mathrm{SO}(3)$ is a rotation and $D[W(\Lambda, \ppp)]$ is its representation.
For spin-$1/2$ particles, the latter is an element of $\mathrm{SU}(2)$, whose
concrete form in terms of momenta and rapidities can be found in
\cite{halpern_special_1968}.

\subsection{Lorentz transformations on particles}

One can now calculate the transformation on the wave function. In the Lorentz
boosted frame, the state is $\ket{\Psi^{\Lambda}} = U(\Lambda) \ket{\Psi}$, so
we have
\begin{align}
\ket{\Psi^{\Lambda}} &= \sum_{\sigma} \int \dddd\mu(p)\, \psi_\sigma(\ppp)
\sum_{\lambda} \ket{\Lambda\ppp, \lambda} D_{\lambda\sigma}[W(\Lambda, \ppp)]
\nonumber\\
&= \sum_{\lambda} \int \dddd\mu(p')\, \sum_{\sigma}
D_{\lambda\sigma}[W(\Lambda, \Lambda^{-1}\ppp')] \psi_\sigma(\Lambda^{-1}\ppp')
\ket{\ppp', \lambda} \nonumber\\
&= \sum_{\lambda} \int \dddd\mu(p)\, \psi^{\Lambda}_{\lambda}(\ppp) \ket{\ppp,
\lambda},
\end{align}
where $\ppp' = \Lambda\ppp$ and we used the fact that the integration measure
is Lorentz covariant, $\dddd\mu(p) = \dddd\mu(\Lambda p)$, with a relabelling
of dummy variables in the last line, $\ppp' \rightarrow \ppp$. Hence we have,
\begin{align}
\psi^{\Lambda}_{\lambda}(\ppp) = \sum_{\sigma} D_{\lambda\sigma}[W(\Lambda,
\Lambda^{-1}\ppp)] \psi_\sigma(\Lambda^{-1}\ppp).
\end{align}

The state of a two particle system belongs to \mbox{$\mathcal{H}_2 =
\mathcal{H}_1 \otimes \mathcal{H}_1$} where $\mathcal{H}_1$ is the one particle
Hilbert space described above. A Lorentz boost $\Lambda$ acts on the two
particle state by $U(\Lambda) \otimes U(\Lambda)$ and in analogy to the single
particle case we calculate that the corresponding transformation of the wave
function is given by
\begin{align}\label{eq:boostedTwoParticleWaveFunction}
\psi^{\Lambda}_{\lambda\kappa}(\ppp, \qqq) &=
\sum_{\sigma, \xi}
D_{\lambda\sigma}\!\left[W(\Lambda, \Lambda^{-1}\ppp)\right]
D_{\kappa\xi}\!\left[W(\Lambda, \Lambda^{-1}\qqq)\right] \nonumber\\
&\phantom{AAAAAAAAA}\times \psi_{\sigma\xi}(\Lambda^{-1}\ppp, \Lambda^{-1}\qqq).
\end{align}

\bibliography{paper_2017_two-continuous-product}

\begin{thebibliography}{46}%
\makeatletter
\providecommand \@ifxundefined [1]{%
 \@ifx{#1\undefined}
}%
\providecommand \@ifnum [1]{%
 \ifnum #1\expandafter \@firstoftwo
 \else \expandafter \@secondoftwo
 \fi
}%
\providecommand \@ifx [1]{%
 \ifx #1\expandafter \@firstoftwo
 \else \expandafter \@secondoftwo
 \fi
}%
\providecommand \natexlab [1]{#1}%
\providecommand \enquote  [1]{``#1''}%
\providecommand \bibnamefont  [1]{#1}%
\providecommand \bibfnamefont [1]{#1}%
\providecommand \citenamefont [1]{#1}%
\providecommand \href@noop [0]{\@secondoftwo}%
\providecommand \href [0]{\begingroup \@sanitize@url \@href}%
\providecommand \@href[1]{\@@startlink{#1}\@@href}%
\providecommand \@@href[1]{\endgroup#1\@@endlink}%
\providecommand \@sanitize@url [0]{\catcode `\\12\catcode `\$12\catcode
  `\&12\catcode `\#12\catcode `\^12\catcode `\_12\catcode `\%12\relax}%
\providecommand \@@startlink[1]{}%
\providecommand \@@endlink[0]{}%
\providecommand \url  [0]{\begingroup\@sanitize@url \@url }%
\providecommand \@url [1]{\endgroup\@href {#1}{\urlprefix }}%
\providecommand \urlprefix  [0]{URL }%
\providecommand \Eprint [0]{\href }%
\providecommand \doibase [0]{http://dx.doi.org/}%
\providecommand \selectlanguage [0]{\@gobble}%
\providecommand \bibinfo  [0]{\@secondoftwo}%
\providecommand \bibfield  [0]{\@secondoftwo}%
\providecommand \translation [1]{[#1]}%
\providecommand \BibitemOpen [0]{}%
\providecommand \bibitemStop [0]{}%
\providecommand \bibitemNoStop [0]{.\EOS\space}%
\providecommand \EOS [0]{\spacefactor3000\relax}%
\providecommand \BibitemShut  [1]{\csname bibitem#1\endcsname}%
\let\auto@bib@innerbib\@empty
\bibitem [{\citenamefont
  {Czachor}(1997)}]{czachor_einstein-podolsky-rosen-bohm_1997}%
  \BibitemOpen
  \bibfield  {author} {\bibinfo {author} {\bibfnamefont {M.}~\bibnamefont
  {Czachor}},\ }\href {\doibase 10.1103/PhysRevA.55.72} {\bibfield  {journal}
  {\bibinfo  {journal} {Physical Review A}\ }\textbf {\bibinfo {volume} {55}},\
  \bibinfo {pages} {72} (\bibinfo {year} {1997})}\BibitemShut {NoStop}%
\bibitem [{\citenamefont {Gingrich}\ and\ \citenamefont
  {Adami}(2002)}]{gingrich_quantum_2002}%
  \BibitemOpen
  \bibfield  {author} {\bibinfo {author} {\bibfnamefont {R.~M.}\ \bibnamefont
  {Gingrich}}\ and\ \bibinfo {author} {\bibfnamefont {C.}~\bibnamefont
  {Adami}},\ }\href {\doibase 10.1103/PhysRevLett.89.270402} {\bibfield
  {journal} {\bibinfo  {journal} {Physical Review Letters}\ }\textbf {\bibinfo
  {volume} {89}},\ \bibinfo {pages} {270402} (\bibinfo {year}
  {2002})}\BibitemShut {NoStop}%
\bibitem [{\citenamefont {Peres}\ \emph {et~al.}(2002)\citenamefont {Peres},
  \citenamefont {Scudo},\ and\ \citenamefont {Terno}}]{peres_quantum_2002}%
  \BibitemOpen
  \bibfield  {author} {\bibinfo {author} {\bibfnamefont {A.}~\bibnamefont
  {Peres}}, \bibinfo {author} {\bibfnamefont {P.~F.}\ \bibnamefont {Scudo}}, \
  and\ \bibinfo {author} {\bibfnamefont {D.~R.}\ \bibnamefont {Terno}},\ }\href
  {\doibase 10.1103/PhysRevLett.88.230402} {\bibfield  {journal} {\bibinfo
  {journal} {Physical Review Letters}\ }\textbf {\bibinfo {volume} {88}},\
  \bibinfo {pages} {230402} (\bibinfo {year} {2002})}\BibitemShut {NoStop}%
\bibitem [{\citenamefont {Alsing}\ and\ \citenamefont
  {Milburn}(2002)}]{alsing_lorentz_2002}%
  \BibitemOpen
  \bibfield  {author} {\bibinfo {author} {\bibfnamefont {P.~M.}\ \bibnamefont
  {Alsing}}\ and\ \bibinfo {author} {\bibfnamefont {G.~J.}\ \bibnamefont
  {Milburn}},\ }\href@noop {} {\bibfield  {journal} {\bibinfo  {journal}
  {Quantum Information and Computation}\ }\textbf {\bibinfo {volume} {2}},\
  \bibinfo {pages} {487} (\bibinfo {year} {2002})}\BibitemShut {NoStop}%
\bibitem [{\citenamefont {Ahn}\ \emph {et~al.}(2002)\citenamefont {Ahn},
  \citenamefont {Lee},\ and\ \citenamefont {Hwang}}]{ahn_relativistic_2002}%
  \BibitemOpen
  \bibfield  {author} {\bibinfo {author} {\bibfnamefont {D.}~\bibnamefont
  {Ahn}}, \bibinfo {author} {\bibfnamefont {H.-J.}\ \bibnamefont {Lee}}, \ and\
  \bibinfo {author} {\bibfnamefont {S.~W.}\ \bibnamefont {Hwang}},\ }\href
  {http://arxiv.org/abs/quant-ph/0207018} {\bibfield  {journal} {\bibinfo
  {journal} {arXiv:quant-ph/0207018}\ } (\bibinfo {year} {2002})}\BibitemShut
  {NoStop}%
\bibitem [{\citenamefont {Ahn}\ \emph {et~al.}(2003{\natexlab{a}})\citenamefont
  {Ahn}, \citenamefont {Lee}, \citenamefont {Moon},\ and\ \citenamefont
  {Hwang}}]{ahn_relativistic_2003}%
  \BibitemOpen
  \bibfield  {author} {\bibinfo {author} {\bibfnamefont {D.}~\bibnamefont
  {Ahn}}, \bibinfo {author} {\bibfnamefont {H.-J.}\ \bibnamefont {Lee}},
  \bibinfo {author} {\bibfnamefont {Y.~H.}\ \bibnamefont {Moon}}, \ and\
  \bibinfo {author} {\bibfnamefont {S.~W.}\ \bibnamefont {Hwang}},\ }\href
  {\doibase 10.1103/PhysRevA.67.012103} {\bibfield  {journal} {\bibinfo
  {journal} {Physical Review A}\ }\textbf {\bibinfo {volume} {67}},\ \bibinfo
  {pages} {012103} (\bibinfo {year} {2003}{\natexlab{a}})}\BibitemShut
  {NoStop}%
\bibitem [{\citenamefont {Ahn}\ \emph {et~al.}(2003{\natexlab{b}})\citenamefont
  {Ahn}, \citenamefont {Lee},\ and\ \citenamefont
  {Hwang}}]{ahn_lorentz-covariant_2003}%
  \BibitemOpen
  \bibfield  {author} {\bibinfo {author} {\bibfnamefont {D.}~\bibnamefont
  {Ahn}}, \bibinfo {author} {\bibfnamefont {H.-J.}\ \bibnamefont {Lee}}, \ and\
  \bibinfo {author} {\bibfnamefont {S.~W.}\ \bibnamefont {Hwang}},\ }\href
  {\doibase 10.1103/PhysRevA.67.032309} {\bibfield  {journal} {\bibinfo
  {journal} {Physical Review A}\ }\textbf {\bibinfo {volume} {67}},\ \bibinfo
  {pages} {032309} (\bibinfo {year} {2003}{\natexlab{b}})}\BibitemShut
  {NoStop}%
\bibitem [{\citenamefont {Czachor}\ and\ \citenamefont
  {Wilczewski}(2003)}]{czachor_relativistic_2003}%
  \BibitemOpen
  \bibfield  {author} {\bibinfo {author} {\bibfnamefont {M.}~\bibnamefont
  {Czachor}}\ and\ \bibinfo {author} {\bibfnamefont {M.}~\bibnamefont
  {Wilczewski}},\ }\href {\doibase 10.1103/PhysRevA.68.010302} {\bibfield
  {journal} {\bibinfo  {journal} {Physical Review A}\ }\textbf {\bibinfo
  {volume} {68}},\ \bibinfo {pages} {010302} (\bibinfo {year}
  {2003})}\BibitemShut {NoStop}%
\bibitem [{\citenamefont {Moon}\ \emph {et~al.}(2004)\citenamefont {Moon},
  \citenamefont {Hwang},\ and\ \citenamefont {Ahn}}]{moon_relativistic_2004}%
  \BibitemOpen
  \bibfield  {author} {\bibinfo {author} {\bibfnamefont {Y.~H.}\ \bibnamefont
  {Moon}}, \bibinfo {author} {\bibfnamefont {S.~W.}\ \bibnamefont {Hwang}}, \
  and\ \bibinfo {author} {\bibfnamefont {D.}~\bibnamefont {Ahn}},\ }\href
  {\doibase 10.1143/PTP.112.219} {\bibfield  {journal} {\bibinfo  {journal}
  {Progress of Theoretical Physics}\ }\textbf {\bibinfo {volume} {112}},\
  \bibinfo {pages} {219} (\bibinfo {year} {2004})}\BibitemShut {NoStop}%
\bibitem [{\citenamefont {Lee}\ and\ \citenamefont
  {Chang-Young}(2004)}]{lee_quantum_2004}%
  \BibitemOpen
  \bibfield  {author} {\bibinfo {author} {\bibfnamefont {D.}~\bibnamefont
  {Lee}}\ and\ \bibinfo {author} {\bibfnamefont {E.}~\bibnamefont
  {Chang-Young}},\ }\href {\doibase 10.1088/1367-2630/6/1/067} {\bibfield
  {journal} {\bibinfo  {journal} {New Journal of Physics}\ }\textbf {\bibinfo
  {volume} {6}},\ \bibinfo {pages} {67} (\bibinfo {year} {2004})}\BibitemShut
  {NoStop}%
\bibitem [{\citenamefont {Czachor}(2005)}]{czachor_comment_2005}%
  \BibitemOpen
  \bibfield  {author} {\bibinfo {author} {\bibfnamefont {M.}~\bibnamefont
  {Czachor}},\ }\href {\doibase 10.1103/PhysRevLett.94.078901} {\bibfield
  {journal} {\bibinfo  {journal} {Physical Review Letters}\ }\textbf {\bibinfo
  {volume} {94}},\ \bibinfo {pages} {078901} (\bibinfo {year}
  {2005})}\BibitemShut {NoStop}%
\bibitem [{\citenamefont {Lamata}\ \emph {et~al.}(2006)\citenamefont {Lamata},
  \citenamefont {Martin-Delgado},\ and\ \citenamefont
  {Solano}}]{lamata_relativity_2006}%
  \BibitemOpen
  \bibfield  {author} {\bibinfo {author} {\bibfnamefont {L.}~\bibnamefont
  {Lamata}}, \bibinfo {author} {\bibfnamefont {M.~A.}\ \bibnamefont
  {Martin-Delgado}}, \ and\ \bibinfo {author} {\bibfnamefont {E.}~\bibnamefont
  {Solano}},\ }\href {\doibase 10.1103/PhysRevLett.97.250502} {\bibfield
  {journal} {\bibinfo  {journal} {Physical Review Letters}\ }\textbf {\bibinfo
  {volume} {97}},\ \bibinfo {pages} {250502} (\bibinfo {year}
  {2006})}\BibitemShut {NoStop}%
\bibitem [{\citenamefont {Alsing}\ \emph {et~al.}(2006)\citenamefont {Alsing},
  \citenamefont {Fuentes-Schuller}, \citenamefont {Mann},\ and\ \citenamefont
  {Tessier}}]{alsing_entanglement_2006}%
  \BibitemOpen
  \bibfield  {author} {\bibinfo {author} {\bibfnamefont {P.~M.}\ \bibnamefont
  {Alsing}}, \bibinfo {author} {\bibfnamefont {I.}~\bibnamefont
  {Fuentes-Schuller}}, \bibinfo {author} {\bibfnamefont {R.~B.}\ \bibnamefont
  {Mann}}, \ and\ \bibinfo {author} {\bibfnamefont {T.~E.}\ \bibnamefont
  {Tessier}},\ }\href {\doibase 10.1103/PhysRevA.74.032326} {\bibfield
  {journal} {\bibinfo  {journal} {Physical Review A}\ }\textbf {\bibinfo
  {volume} {74}},\ \bibinfo {pages} {032326} (\bibinfo {year}
  {2006})}\BibitemShut {NoStop}%
\bibitem [{\citenamefont {Chakrabarti}(2009)}]{chakrabarti_entangled_2009}%
  \BibitemOpen
  \bibfield  {author} {\bibinfo {author} {\bibfnamefont {A.}~\bibnamefont
  {Chakrabarti}},\ }\href {\doibase 10.1088/1751-8113/42/24/245205} {\bibfield
  {journal} {\bibinfo  {journal} {Journal of Physics A: Mathematical and
  Theoretical}\ }\textbf {\bibinfo {volume} {42}},\ \bibinfo {pages} {245205}
  (\bibinfo {year} {2009})}\BibitemShut {NoStop}%
\bibitem [{\citenamefont {Fuentes}\ \emph {et~al.}(2010)\citenamefont
  {Fuentes}, \citenamefont {Mann}, \citenamefont {Mart{\'i}n-Mart{\'i}nez},\
  and\ \citenamefont {Moradi}}]{fuentes_entanglement_2010}%
  \BibitemOpen
  \bibfield  {author} {\bibinfo {author} {\bibfnamefont {I.}~\bibnamefont
  {Fuentes}}, \bibinfo {author} {\bibfnamefont {R.~B.}\ \bibnamefont {Mann}},
  \bibinfo {author} {\bibfnamefont {E.}~\bibnamefont
  {Mart{\'i}n-Mart{\'i}nez}}, \ and\ \bibinfo {author} {\bibfnamefont
  {S.}~\bibnamefont {Moradi}},\ }\href {\doibase 10.1103/PhysRevD.82.045030}
  {\bibfield  {journal} {\bibinfo  {journal} {Physical Review D}\ }\textbf
  {\bibinfo {volume} {82}},\ \bibinfo {pages} {045030} (\bibinfo {year}
  {2010})}\BibitemShut {NoStop}%
\bibitem [{\citenamefont {Friis}\ \emph {et~al.}(2010)\citenamefont {Friis},
  \citenamefont {Bertlmann},\ and\ \citenamefont
  {Huber}}]{friis_relativistic_2010}%
  \BibitemOpen
  \bibfield  {author} {\bibinfo {author} {\bibfnamefont {N.}~\bibnamefont
  {Friis}}, \bibinfo {author} {\bibfnamefont {R.~A.}\ \bibnamefont
  {Bertlmann}}, \ and\ \bibinfo {author} {\bibfnamefont {M.}~\bibnamefont
  {Huber}},\ }\href {\doibase 10.1103/PhysRevA.81.042114} {\bibfield  {journal}
  {\bibinfo  {journal} {Physical Review A}\ }\textbf {\bibinfo {volume} {81}},\
  \bibinfo {pages} {042114} (\bibinfo {year} {2010})}\BibitemShut {NoStop}%
\bibitem [{\citenamefont {Palge}\ and\ \citenamefont
  {Dunningham}(2012)}]{palge_generation_2012}%
  \BibitemOpen
  \bibfield  {author} {\bibinfo {author} {\bibfnamefont {V.}~\bibnamefont
  {Palge}}\ and\ \bibinfo {author} {\bibfnamefont {J.}~\bibnamefont
  {Dunningham}},\ }\href {\doibase 10.1103/PhysRevA.85.042322} {\bibfield
  {journal} {\bibinfo  {journal} {Physical Review A}\ }\textbf {\bibinfo
  {volume} {85}},\ \bibinfo {pages} {042322} (\bibinfo {year}
  {2012})}\BibitemShut {NoStop}%
\bibitem [{\citenamefont {Alsing}\ and\ \citenamefont
  {Fuentes}(2012)}]{alsing_observer-dependent_2012}%
  \BibitemOpen
  \bibfield  {author} {\bibinfo {author} {\bibfnamefont {P.~M.}\ \bibnamefont
  {Alsing}}\ and\ \bibinfo {author} {\bibfnamefont {I.}~\bibnamefont
  {Fuentes}},\ }\href {\doibase 10.1088/0264-9381/29/22/224001} {\bibfield
  {journal} {\bibinfo  {journal} {Classical and Quantum Gravity}\ }\textbf
  {\bibinfo {volume} {29}},\ \bibinfo {pages} {224001} (\bibinfo {year}
  {2012})}\BibitemShut {NoStop}%
\bibitem [{\citenamefont {Terashima}\ and\ \citenamefont
  {Ueda}(2003)}]{terashima_relativistic_2003}%
  \BibitemOpen
  \bibfield  {author} {\bibinfo {author} {\bibfnamefont {H.}~\bibnamefont
  {Terashima}}\ and\ \bibinfo {author} {\bibfnamefont {M.}~\bibnamefont
  {Ueda}},\ }\href {\doibase 10.1142/S0219749903000061} {\bibfield  {journal}
  {\bibinfo  {journal} {International Journal of Quantum Information}\ }\textbf
  {\bibinfo {volume} {01}},\ \bibinfo {pages} {93} (\bibinfo {year}
  {2003})}\BibitemShut {NoStop}%
\bibitem [{\citenamefont {Caban}\ and\ \citenamefont
  {Rembieli{\'n}ski}(2005)}]{caban_lorentz-covariant_2005}%
  \BibitemOpen
  \bibfield  {author} {\bibinfo {author} {\bibfnamefont {P.}~\bibnamefont
  {Caban}}\ and\ \bibinfo {author} {\bibfnamefont {J.}~\bibnamefont
  {Rembieli{\'n}ski}},\ }\href {\doibase 10.1103/PhysRevA.72.012103} {\bibfield
   {journal} {\bibinfo  {journal} {Physical Review A}\ }\textbf {\bibinfo
  {volume} {72}},\ \bibinfo {pages} {012103} (\bibinfo {year}
  {2005})}\BibitemShut {NoStop}%
\bibitem [{\citenamefont {Caban}\ and\ \citenamefont
  {Rembieli{\'n}ski}(2006)}]{caban_einstein-podolsky-rosen_2006}%
  \BibitemOpen
  \bibfield  {author} {\bibinfo {author} {\bibfnamefont {P.}~\bibnamefont
  {Caban}}\ and\ \bibinfo {author} {\bibfnamefont {J.}~\bibnamefont
  {Rembieli{\'n}ski}},\ }\href {\doibase 10.1103/PhysRevA.74.042103} {\bibfield
   {journal} {\bibinfo  {journal} {Physical Review A}\ }\textbf {\bibinfo
  {volume} {74}},\ \bibinfo {pages} {042103} (\bibinfo {year}
  {2006})}\BibitemShut {NoStop}%
\bibitem [{\citenamefont {Jordan}\ \emph {et~al.}(2006)\citenamefont {Jordan},
  \citenamefont {Shaji},\ and\ \citenamefont {Sudarshan}}]{jordan_maps_2006}%
  \BibitemOpen
  \bibfield  {author} {\bibinfo {author} {\bibfnamefont {T.}~\bibnamefont
  {Jordan}}, \bibinfo {author} {\bibfnamefont {A.}~\bibnamefont {Shaji}}, \
  and\ \bibinfo {author} {\bibfnamefont {E.}~\bibnamefont {Sudarshan}},\ }\href
  {\doibase 10.1103/PhysRevA.73.032104} {\bibfield  {journal} {\bibinfo
  {journal} {Physical Review A}\ }\textbf {\bibinfo {volume} {73}},\ \bibinfo
  {pages} {032104} (\bibinfo {year} {2006})}\BibitemShut {NoStop}%
\bibitem [{\citenamefont {Jordan}\ \emph {et~al.}(2007)\citenamefont {Jordan},
  \citenamefont {Shaji},\ and\ \citenamefont
  {Sudarshan}}]{jordan_lorentz_2007}%
  \BibitemOpen
  \bibfield  {author} {\bibinfo {author} {\bibfnamefont {T.}~\bibnamefont
  {Jordan}}, \bibinfo {author} {\bibfnamefont {A.}~\bibnamefont {Shaji}}, \
  and\ \bibinfo {author} {\bibfnamefont {E.}~\bibnamefont {Sudarshan}},\ }\href
  {\doibase 10.1103/PhysRevA.75.022101} {\bibfield  {journal} {\bibinfo
  {journal} {Physical Review A}\ }\textbf {\bibinfo {volume} {75}},\ \bibinfo
  {pages} {022101} (\bibinfo {year} {2007})}\BibitemShut {NoStop}%
\bibitem [{\citenamefont {Palge}\ \emph {et~al.}(2011)\citenamefont {Palge},
  \citenamefont {Vedral},\ and\ \citenamefont
  {Dunningham}}]{palge_behavior_2011}%
  \BibitemOpen
  \bibfield  {author} {\bibinfo {author} {\bibfnamefont {V.}~\bibnamefont
  {Palge}}, \bibinfo {author} {\bibfnamefont {V.}~\bibnamefont {Vedral}}, \
  and\ \bibinfo {author} {\bibfnamefont {J.~A.}\ \bibnamefont {Dunningham}},\
  }\href {\doibase 10.1103/PhysRevA.84.044303} {\bibfield  {journal} {\bibinfo
  {journal} {Physical Review A}\ }\textbf {\bibinfo {volume} {84}},\ \bibinfo
  {pages} {044303} (\bibinfo {year} {2011})}\BibitemShut {NoStop}%
\bibitem [{Note1()}]{Note1}%
  \BibitemOpen
  \bibinfo {note} {A related study that involves systems with discrete momenta
  along with an extension to mixed spin states can be found in \cite
  {palge_behavior_2015}.}\BibitemShut {Stop}%
\bibitem [{Note2()}]{Note2}%
  \BibitemOpen
  \bibinfo {note} {The unitary irreducible representation is due to Wigner
  \cite {wigner_unitary_1939}. A standard treatment of the two representations
  is given by \cite {bogolubov_introduction_1975}. See \cite
  {polyzou_spin_2012} for a very readable account on the relationship between
  the two frameworks and \cite {caban_spin_2013} for a discussion in the
  context of spin observable in the Dirac theory.}\BibitemShut {Stop}%
\bibitem [{\citenamefont {Caban}\ \emph {et~al.}(2013)\citenamefont {Caban},
  \citenamefont {Rembieli{\'n}ski},\ and\ \citenamefont
  {W{\l}odarczyk}}]{caban_spin_2013}%
  \BibitemOpen
  \bibfield  {author} {\bibinfo {author} {\bibfnamefont {P.}~\bibnamefont
  {Caban}}, \bibinfo {author} {\bibfnamefont {J.}~\bibnamefont
  {Rembieli{\'n}ski}}, \ and\ \bibinfo {author} {\bibfnamefont
  {M.}~\bibnamefont {W{\l}odarczyk}},\ }\href
  {http://journals.aps.org/pra/abstract/10.1103/PhysRevA.88.022119} {\bibfield
  {journal} {\bibinfo  {journal} {Physical Review A}\ }\textbf {\bibinfo
  {volume} {88}},\ \bibinfo {pages} {022119} (\bibinfo {year}
  {2013})}\BibitemShut {NoStop}%
\bibitem [{Note3()}]{Note3}%
  \BibitemOpen
  \bibinfo {note} {We will use the abbreviation TWR in honor of Thomas's
  contribution of discovering the Thomas precession \cite
  {thomas_motion_1926,thomas_kinematics_1927}.}\BibitemShut {Stop}%
\bibitem [{Note4()}]{Note4}%
  \BibitemOpen
  \bibinfo {note} {With the caveat that strictly considered the momentum state
  is not a qubit, so we call it a momentum system.}\BibitemShut {Stop}%
\bibitem [{Note5()}]{Note5}%
  \BibitemOpen
  \bibinfo {note} {Entangled momenta are discussed in a related
  paper.}\BibitemShut {Stop}%
\bibitem [{\citenamefont {Palge}\ and\ \citenamefont
  {Dunningham}(2015)}]{palge_behavior_2015}%
  \BibitemOpen
  \bibfield  {author} {\bibinfo {author} {\bibfnamefont {V.}~\bibnamefont
  {Palge}}\ and\ \bibinfo {author} {\bibfnamefont {J.}~\bibnamefont
  {Dunningham}},\ }\href {\doibase 10.1016/j.aop.2015.09.028} {\bibfield
  {journal} {\bibinfo  {journal} {Annals of Physics}\ }\textbf {\bibinfo
  {volume} {363}},\ \bibinfo {pages} {275} (\bibinfo {year}
  {2015})}\BibitemShut {NoStop}%
\bibitem [{\citenamefont {Sexl}\ and\ \citenamefont
  {Urbantke}(2001)}]{sexl_relativity_2001}%
  \BibitemOpen
  \bibfield  {author} {\bibinfo {author} {\bibfnamefont {R.~U.}\ \bibnamefont
  {Sexl}}\ and\ \bibinfo {author} {\bibfnamefont {H.~K.}\ \bibnamefont
  {Urbantke}},\ }\href@noop {} {\emph {\bibinfo {title} {Relativity, {Groups},
  {Particles}: {Special} {Relativity} and {Relativistic} {Symmetry} in {Field}
  and {Particle} {Physics}}}},\ \bibinfo {edition} {rev. ed.}\ ed.\ (\bibinfo
  {publisher} {Springer},\ \bibinfo {address} {New York},\ \bibinfo {year}
  {2001})\BibitemShut {NoStop}%
\bibitem [{\citenamefont {Bogolubov}\ \emph {et~al.}(1975)\citenamefont
  {Bogolubov}, \citenamefont {Logunov},\ and\ \citenamefont
  {Todorov}}]{bogolubov_introduction_1975}%
  \BibitemOpen
  \bibfield  {author} {\bibinfo {author} {\bibfnamefont {N.~N.}\ \bibnamefont
  {Bogolubov}}, \bibinfo {author} {\bibfnamefont {A.~A.}\ \bibnamefont
  {Logunov}}, \ and\ \bibinfo {author} {\bibfnamefont {I.~T.}\ \bibnamefont
  {Todorov}},\ }\href@noop {} {\emph {\bibinfo {title} {Introduction to
  {Axiomatic} {Quantum} {Field} {Theory}}}}\ (\bibinfo  {publisher}
  {W.A.Benjamin},\ \bibinfo {year} {1975})\BibitemShut {NoStop}%
\bibitem [{\citenamefont {MacFarlane}(1963)}]{macfarlane_kinematics_1963}%
  \BibitemOpen
  \bibfield  {author} {\bibinfo {author} {\bibfnamefont {A.~J.}\ \bibnamefont
  {MacFarlane}},\ }\href {\doibase 10.1063/1.1703981} {\bibfield  {journal}
  {\bibinfo  {journal} {Journal of Mathematical Physics}\ }\textbf {\bibinfo
  {volume} {4}},\ \bibinfo {pages} {490} (\bibinfo {year} {1963})}\BibitemShut
  {NoStop}%
\bibitem [{\citenamefont {Wootters}(1998)}]{wootters_entanglement_1998}%
  \BibitemOpen
  \bibfield  {author} {\bibinfo {author} {\bibfnamefont {W.}~\bibnamefont
  {Wootters}},\ }\href {\doibase 10.1103/PhysRevLett.80.2245} {\bibfield
  {journal} {\bibinfo  {journal} {Physical Review Letters}\ }\textbf {\bibinfo
  {volume} {80}},\ \bibinfo {pages} {2245} (\bibinfo {year}
  {1998})}\BibitemShut {NoStop}%
\bibitem [{\citenamefont {Rhodes}\ and\ \citenamefont
  {Semon}(2004)}]{rhodes_relativistic_2004}%
  \BibitemOpen
  \bibfield  {author} {\bibinfo {author} {\bibfnamefont {J.~A.}\ \bibnamefont
  {Rhodes}}\ and\ \bibinfo {author} {\bibfnamefont {M.~D.}\ \bibnamefont
  {Semon}},\ }\href {\doibase 10.1119/1.1652040} {\bibfield  {journal}
  {\bibinfo  {journal} {Am. J. Phys.}\ }\textbf {\bibinfo {volume} {72}},\
  \bibinfo {pages} {943} (\bibinfo {year} {2004})}\BibitemShut {NoStop}%
\bibitem [{\citenamefont {Halpern}(1968)}]{halpern_special_1968}%
  \BibitemOpen
  \bibfield  {author} {\bibinfo {author} {\bibfnamefont {F.~R.}\ \bibnamefont
  {Halpern}},\ }\href@noop {} {\emph {\bibinfo {title} {Special {Relativity}
  and {Quantum} {Mechanics}}}}\ (\bibinfo  {publisher} {Prentice-Hall},\
  \bibinfo {year} {1968})\BibitemShut {NoStop}%
\bibitem [{\citenamefont {Einstein}\ \emph {et~al.}(1935)\citenamefont
  {Einstein}, \citenamefont {Podolsky},\ and\ \citenamefont
  {Rosen}}]{einstein_can_1935}%
  \BibitemOpen
  \bibfield  {author} {\bibinfo {author} {\bibfnamefont {A.}~\bibnamefont
  {Einstein}}, \bibinfo {author} {\bibfnamefont {B.}~\bibnamefont {Podolsky}},
  \ and\ \bibinfo {author} {\bibfnamefont {N.}~\bibnamefont {Rosen}},\ }\href
  {\doibase 10.1103/PhysRev.47.777} {\bibfield  {journal} {\bibinfo  {journal}
  {Physical Review}\ }\textbf {\bibinfo {volume} {47}},\ \bibinfo {pages} {777}
  (\bibinfo {year} {1935})}\BibitemShut {NoStop}%
\bibitem [{\citenamefont {Bohm}(1951)}]{bohm_quantum_1951}%
  \BibitemOpen
  \bibfield  {author} {\bibinfo {author} {\bibfnamefont {D.}~\bibnamefont
  {Bohm}},\ }\href@noop {} {\emph {\bibinfo {title} {Quantum {Theory}}}},\
  Prentice-{Hall} physics series\ (\bibinfo  {publisher} {Prentice-Hall},\
  \bibinfo {address} {New York},\ \bibinfo {year} {1951})\BibitemShut {NoStop}%
\bibitem [{Note6()}]{Note6}%
  \BibitemOpen
  \bibinfo {note} {From a more general perspective, it is interesting to
  consider mixed states as well. See \cite {palge_behavior_2015} for an
  extension to the Werner states.}\BibitemShut {Stop}%
\bibitem [{\citenamefont {Bengtsson}\ and\ \citenamefont
  {{\.Z}yczkowski}(2006)}]{bengtsson_geometry_2006}%
  \BibitemOpen
  \bibfield  {author} {\bibinfo {author} {\bibfnamefont {I.}~\bibnamefont
  {Bengtsson}}\ and\ \bibinfo {author} {\bibfnamefont {K.}~\bibnamefont
  {{\.Z}yczkowski}},\ }\href@noop {} {\emph {\bibinfo {title} {Geometry of
  {Quantum} {States}: {An} {Introduction} to {Quantum} {Entanglement}}}}\
  (\bibinfo  {publisher} {Cambridge University Press},\ \bibinfo {year}
  {2006})\BibitemShut {NoStop}%
\bibitem [{\citenamefont {Bertlmann}\ \emph {et~al.}(2002)\citenamefont
  {Bertlmann}, \citenamefont {Narnhofer},\ and\ \citenamefont
  {Thirring}}]{bertlmann_geometric_2002}%
  \BibitemOpen
  \bibfield  {author} {\bibinfo {author} {\bibfnamefont {R.~A.}\ \bibnamefont
  {Bertlmann}}, \bibinfo {author} {\bibfnamefont {H.}~\bibnamefont
  {Narnhofer}}, \ and\ \bibinfo {author} {\bibfnamefont {W.}~\bibnamefont
  {Thirring}},\ }\href {\doibase 10.1103/PhysRevA.66.032319} {\bibfield
  {journal} {\bibinfo  {journal} {Physical Review A}\ }\textbf {\bibinfo
  {volume} {66}},\ \bibinfo {pages} {032319} (\bibinfo {year}
  {2002})}\BibitemShut {NoStop}%
\bibitem [{\citenamefont {Wigner}(1939)}]{wigner_unitary_1939}%
  \BibitemOpen
  \bibfield  {author} {\bibinfo {author} {\bibfnamefont {E.}~\bibnamefont
  {Wigner}},\ }\href {http://www.jstor.org/stable/1968551} {\bibfield
  {journal} {\bibinfo  {journal} {Annals of Mathematics}\ }\bibinfo {series}
  {Second {Series}},\ \textbf {\bibinfo {volume} {40}},\ \bibinfo {pages} {149}
  (\bibinfo {year} {1939})}\BibitemShut {NoStop}%
\bibitem [{\citenamefont {Polyzou}\ \emph {et~al.}(2012)\citenamefont
  {Polyzou}, \citenamefont {Gl{\"o}ckle},\ and\ \citenamefont
  {Wita{\l}a}}]{polyzou_spin_2012}%
  \BibitemOpen
  \bibfield  {author} {\bibinfo {author} {\bibfnamefont {W.~N.}\ \bibnamefont
  {Polyzou}}, \bibinfo {author} {\bibfnamefont {W.}~\bibnamefont
  {Gl{\"o}ckle}}, \ and\ \bibinfo {author} {\bibfnamefont {H.}~\bibnamefont
  {Wita{\l}a}},\ }\href {\doibase 10.1007/s00601-012-0526-8} {\bibfield
  {journal} {\bibinfo  {journal} {Few-Body Systems}\ }\textbf {\bibinfo
  {volume} {54}},\ \bibinfo {pages} {1667} (\bibinfo {year}
  {2012})}\BibitemShut {NoStop}%
\bibitem [{\citenamefont {Thomas}(1926)}]{thomas_motion_1926}%
  \BibitemOpen
  \bibfield  {author} {\bibinfo {author} {\bibfnamefont {L.~H.}\ \bibnamefont
  {Thomas}},\ }\href {\doibase 10.1038/117514a0} {\bibfield  {journal}
  {\bibinfo  {journal} {Nature}\ }\textbf {\bibinfo {volume} {117}},\ \bibinfo
  {pages} {514} (\bibinfo {year} {1926})}\BibitemShut {NoStop}%
\bibitem [{\citenamefont {Thomas}(1927)}]{thomas_kinematics_1927}%
  \BibitemOpen
  \bibfield  {author} {\bibinfo {author} {\bibfnamefont {L.~H.}\ \bibnamefont
  {Thomas}},\ }\href {\doibase 10.1080/14786440108564170} {\bibfield  {journal}
  {\bibinfo  {journal} {Philosophical Magazine}\ }\bibinfo {series} {7},\
  \textbf {\bibinfo {volume} {3}},\ \bibinfo {pages} {1} (\bibinfo {year}
  {1927})}\BibitemShut {NoStop}%
\end{thebibliography}%

\end{document}